\documentclass[acmtog]{acmart}

\usepackage{booktabs} 
\usepackage{overpic}
\usepackage{wrapfig}
\usepackage{multirow} 
\usepackage{soul}

\newcommand{\under}[1]{\underline{\textbf{#1}}}

\usepackage[ruled]{algorithm2e} 

\SetAlFnt{\small}
\SetAlCapFnt{\small}
\SetAlCapNameFnt{\small}
\SetAlCapHSkip{0pt}

\acmJournal{TOG}

\begin{document}
\title{CrossGen: Learning and Generating Cross Fields for Quad Meshing}

\setcopyright{cc}
\setcctype{by}
\acmJournal{TOG}
\acmYear{2025} \acmVolume{44} \acmNumber{6} \acmArticle{} \acmMonth{12} \acmPrice{}\acmDOI{10.1145/3763299}

\author{Qiujie Dong}
\authornote{Equal contribution}
    \orcid{0000-0001-6271-2546}
    \affiliation{%
    \institution{The University of Hong Kong}
    \city{Hong Kong}
    \country{China}}
    \affiliation{%
    \institution{Shandong University}
    \city{Qingdao}
    \country{China}}
    \email{qiujie.jay.dong@gmail.com}

\author{Jiepeng Wang}
\authornotemark[1]
\authornote{Corresponding authors.}
    \orcid{0000-0002-6049-4458}
    \affiliation{%
    \institution{The University of Hong Kong}
    \city{Hong Kong}
    \country{China}}
    \email{jiepeng@connect.hku.hk}
    
\author{Rui Xu}
    \orcid{0000-0001-8273-1808}
    \affiliation{%
    \institution{The University of Hong Kong}
    \city{Hong Kong}
    \country{China}}
    \email{ruixu1999@connect.hku.hk}
    
\author{Cheng Lin}
    \orcid{0000-0002-3335-6623}
    \affiliation{%
    \institution{Macau University of Science and Technology}
    \city{Macau}
    \country{China}}
    \email{chlin@connect.hku.hk}

\author{Yuan Liu}
    \orcid{0000-0003-2933-5667}
    \affiliation{%
    \institution{Hong Kong University of Science and Technology}
    \city{Hong Kong}
    \country{China}}
    \email{liuyuanwhuer@gmail.com}
    
\author{Shiqing Xin}      
    \orcid{0000-0001-8452-8723}
    \affiliation{%
    \institution{Shandong University}
    \city{Qingdao}
    \country{China}}
    \email{xinshiqing@sdu.edu.cn}

\author{Zichun Zhong}      
    \orcid{0000-0001-6489-6502}
    \affiliation{%
    \institution{Wayne State University}
    \city{Detroit}
    \country{United States of America}}
    \email{zichunzhong@wayne.edu}

\author{Xin Li}      
    \orcid{0000-0002-0144-9489}
    \affiliation{%
    \institution{Texas A\&M University}
    \city{Texas}
    \country{United States of America}}
    \email{xinli@tamu.edu}
    
\author{Changhe Tu}
    \orcid{0000-0002-1231-3392}
    \affiliation{%
    \institution{Shandong University}
    \city{Qingdao}
    \country{China}}
    \email{chtu@sdu.edu.cn}

\author{Taku Komura}
    \orcid{0000-0002-2729-5860}
    \affiliation{%
    \institution{The University of Hong Kong}
    \city{Hong Kong}
    \country{China}}
    \email{taku@cs.hku.hk}

\author{Leif Kobbelt}
    \orcid{0000-0002-7880-9470}
    \affiliation{%
    \institution{RWTH Aachen University}
    \city{Aachen}
    \country{Germany}}
    \email{sekretariati8@informatik.rwth-aachen.de}

\author{Scott Schaefer}
    \orcid{0000-0002-0988-1452}
    \affiliation{%
    \institution{Texas A\&M University}
    \city{Texas}
    \country{United States of America}}
    \email{schaefer@cse.tamu.edu}

\author{Wenping Wang}
\authornotemark[2]
    \orcid{0000-0002-2284-3952}
    \affiliation{%
    \institution{Texas A\&M University}
    \state{Texas}
    \country{United States of America}}
    \email{wenping@tamu.edu}

\begin{abstract}
Cross fields play a critical role in various geometry processing tasks, especially for quad mesh generation.
Existing methods for cross field generation often struggle to balance computational efficiency with generation quality, using slow per-shape optimization.
We introduce \emph{CrossGen}, a novel framework that supports both feed-forward prediction and latent generative modeling of cross fields for quad meshing by unifying geometry and cross field representations within a joint latent space. 
Our method enables extremely fast computation of high-quality cross fields of general input shapes, typically within one second
without per-shape optimization. 
Our method assumes a point-sampled surface, also called a {\em point-cloud surface}, as input, so we can accommodate various surface representations by a straightforward point sampling process. 
Using an auto-encoder network architecture, we encode input point-cloud surfaces
into a sparse voxel grid with fine-grained latent spaces, which are decoded into both SDF-based surface geometry and cross fields~(see the teaser figure). 
We also contribute a dataset of models with both high-quality signed distance fields (SDFs) representations and their corresponding cross fields, and use it to train our network. 
Once trained, the network is capable of computing a cross field of an input surface in a feed-forward manner, ensuring high geometric fidelity, noise resilience, and rapid inference.
Furthermore, leveraging the same unified latent representation, we incorporate a diffusion model for computing cross fields of new shapes generated from partial input, such as sketches. 
To demonstrate its practical applications, we validate \emph{CrossGen} on the quad mesh generation task for a large variety of surface shapes.
Experimental results demonstrate that \emph{CrossGen} generalizes well across diverse shapes and consistently yields high-fidelity cross fields, thus facilitating the generation of high-quality quad meshes.
\end{abstract}

%
%


%
%


\begin{teaserfigure}
    \centering
    \vspace{-1mm}
    \includegraphics[width=\textwidth]{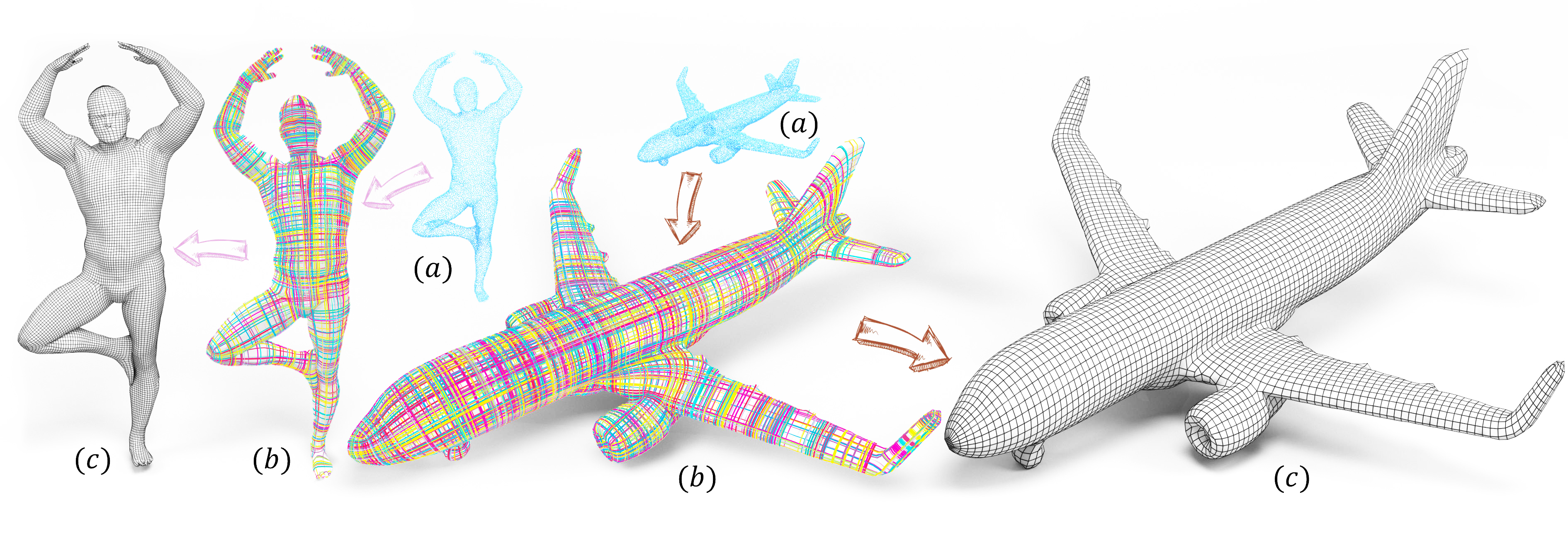}
    \vspace{-10mm}
    \caption{
   Results produced by our method, \emph{CrossGen}, 
   that predicts cross fields on given shapes for quad meshing in a feed-forward manner without per shape optimization. (a) Input shapes as point cloud surfaces; (b) Cross fields generated by    \emph{CrossGen}; and (c) The resulting quad meshes.   
   \emph{CrossGen} demonstrates significant advantages in terms of efficiency and generalizability across various shape types, enabling fast, high-quality quad mesh generation for downstream applications. 
    }
    \label{fig:teaser}
\end{teaserfigure}

\maketitle

\section{Introduction}\label{sec:intro}
Cross fields are fundamental in various geometry processing tasks, such as quadrilateral (quad) mesh generation~\cite{DL2quadMesh2021, IGM2013, Justin2020, viertel2019approach, beaufort2017computing, brandt2018modeling, palmer2024lifting}, surface subdivision~\cite{CC_subdivision2008, subdivision2014}, layout generation \cite{campen2012dualloopmeshing}, and shape parameterization (T-spline design)~\cite{TSpline2017}, as they provide a powerful way to capture the underlying structure and flow of surface geometry.
By aligning with principal curvature directions and conforming to sharp feature edges or the boundaries of open surfaces, cross fields effectively capture the natural flow of the underlying geometry, leading to more coherent and visually consistent results.

\begin{figure}[t]
  \centering
   \begin{overpic}[width=\linewidth]{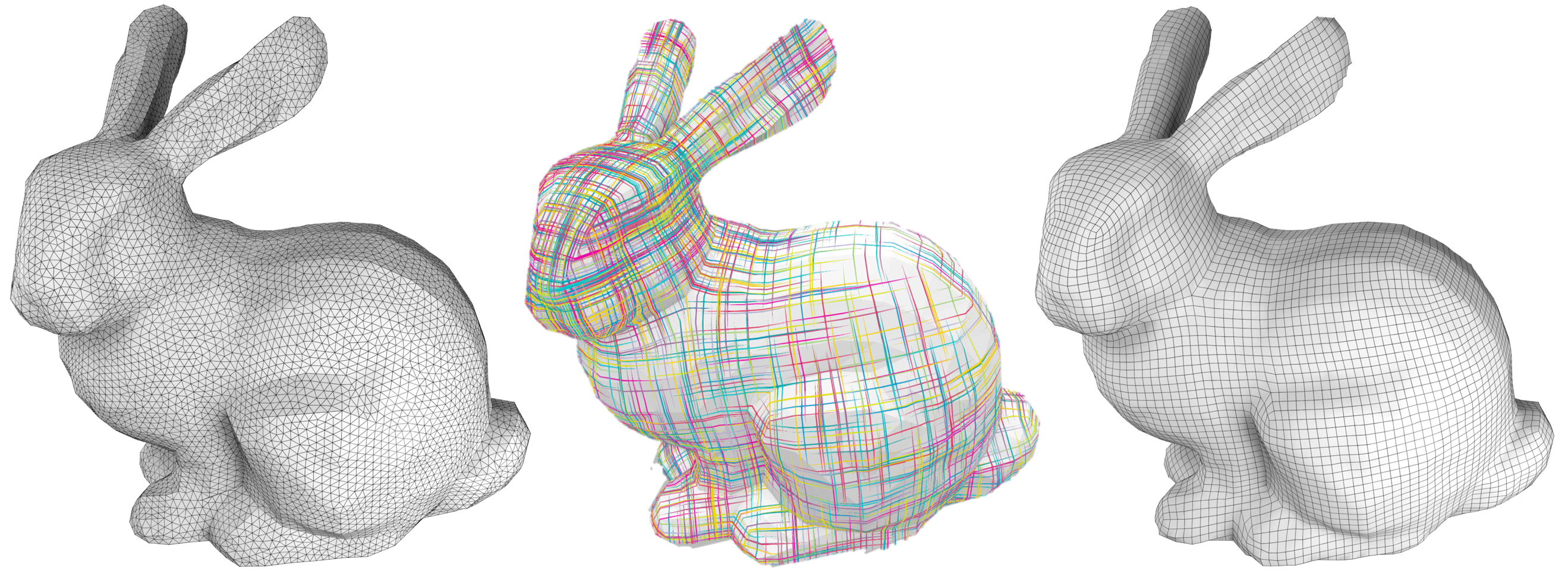}
        \put(9,-3){\small (a) Input mesh}
        \put(42,-3){\small (b) Cross field}
        \put(75,-3){\small (c) Quad mesh}
   \end{overpic}
   \caption{A standard pipeline of quad meshing based on cross fields. Given a triangle mesh (a), a cross field (b) is computed on the surface, and then a quadrilateral mesh (c) is extracted with the guidance of the cross field.}
   \label{fig:vis_cf2quad}
\end{figure}

The main driving force for cross field generation is the task of quad meshing, which plays a pivotal role in computer graphics, engineering, and scientific computing. From character animation and digital sculpting to physics simulation and architectural geometry, quad meshes offer benefits such as simpler subdivision schemes and compatibility with downstream applications such as parameterization and spline construction. A major factor determining the fidelity and adaptability of a quad mesh is the \emph{cross field} (see Fig. \ref{fig:vis_cf2quad}). It is noted that better fields can produce quad meshes with fewer distortions, more uniform spacing, and improved alignment with intrinsic surface features~\cite{fieldDesign2016, quadwild2021}.

Although high-quality cross fields offer clear advantages, generating cross fields that accurately capture the intrinsic geometry of complex shapes is non-trivial. Most methods rely on slow per-shape optimization, such as~\citet{NeurCross2024} and ~\citet{QuadriFlow2018}, which can take several minutes for a single shape. These methods do not generalize to new shapes without optimization. 
However, data-driven approaches for generalizable cross field generation remain under-explored. One closely related work is~\citet{DL2quadMesh2021}, which proposes a method specifically for learning directional fields for the human mesh category using a data-driven approach. This method is specifically designed and tested on a single category and relies on global shape encoding. 
Extending it to generalizable cross field generation across various shape categories is an open problem. 
These optimization-based methods or data-driven approaches are typically slow or lack demonstrated generalization ability across categories, limiting their practical utility. 

In this work, we aim to address the above challenges by focusing on two core objectives:
(1) \emph{Fast high-fidelity cross field generation}, so 
cross fields can be computed in milliseconds without the need for slow iterative per-shape optimization; and
(2) \emph{Generalization across a broad range of shape types}, including smooth organic surfaces, common objects, CAD models, and surfaces with open boundaries.

To fulfill these objectives, we propose a novel framework, called \emph{CrossGen}. 
The input to our method is assumed to a point-cloud surface, which can be a raw point cloud or points sampled from another surface representation, such as a triangle mesh surface. Hence we may also accommodate an input shape given in any surface representation since it can be converted to a point-cloud surface through sampling. 
Furthermore, we assume that each point of the input is associated with a surface normal vector, 
which is either provided directly or estimated using the method proposed by \citet{normal_Xu506}.
Our network simultaneously outputs a signed distance field (SDF) of the input shape and the cross field. The zero-level set of the SDF provides a continuous surface for extracting a quad mesh as guided by the output cross field.
See the teaser figure for an illustration.

\emph{CrossGen} adopts an auto-encoder architecture. The encoder transforms the input point clouds into a latent space composed of sparse grids of embeddings. These embeddings are then decoded into a high-resolution feature grid, which supports the simultaneous prediction of the SDF and cross field through two multi-layer perception~(MLP) branches.
Note that the encoder predominantly captures localized geometric details rather than global shape semantics, in order to enhance cross-categorical generalization capacity across diverse topological classes. 
Additionally, leveraging the same learned latent space, we explore the extension of our framework to generative modeling by incorporating a two-stage diffusion pipeline in the latent space. With this extension, the synthesis of novel 3D shapes and the computation of their cross fields are carried out in an end-to-end manner.

To train our network, we need a dataset of diverse shapes with annotations of paired SDF-based surfaces and high-quality cross fields.
To the best of our knowledge, there is currently no publicly available dataset of this kind.
Hence, we have made a diverse dataset of 3D shapes. 
For each shape, we compute an SDF to capture its geometry, along with a corresponding cross field aligned with principal curvature directions or sharp features. 
The resulting dataset comprises over 10,000 shapes, offering rich geometric and structural diversity to support robust and generalizable model training.

In summary, we make the following contributions:

\begin{itemize} 
\item \textbf{Unified geometry representation.} We develop a novel auto-encoder network as the backbone of \emph{CrossGen}
   for robust learning of cross-fields of a large variety of input shapes. Specifically, the unique design of the encoder with a local perception field enables \emph{CrossGen} to generalize well across diverse shape types and remain robust to out-of-domain inputs and rotational variations. For inference, \emph{CrossGen} computes a cross-field within one second, several orders of magnitude faster over the state-of-the-art optimization-based methods. 

\item \textbf{Large-scale dataset.} We contribute a training dataset of over 10,000 shapes annotated with high-quality SDFs and cross field, which is the first dataset of its kind. 

\item \textbf{Performance validation.} We present extensive validation and comparisons of \emph{CrossGen} with the existing methods to demonstrate the efficacy of \emph{CrossGen} in terms of efficiency and quality of the cross fields and quad meshes computed.

\end{itemize}

We will publicly release both our dataset and code. In addition to high-quality SDFs and cross field pairs, the dataset will also include quad meshes extracted from the cross fields, facilitating further research and downstream applications.

\section{Related Work}
\subsection{Cross Field Generation}\label{sec:cf_related_works_gen}
There are numerous existing works~\cite{Chebyshev2019, Integrable2015, Panozzo2014, N-symmetry2008, Illustrating2000, survery2017} on generating cross fields of polygonal mesh surfaces. These efforts can be classified into two categories: optimization-based methods and data-driven methods.

\paragraph{Optimization-based Methods}
Several methods utilize optimization to create cross fields.
Mixed-Integer Quadrangulation~(MIQ) \cite{MIQ2009} initializes the cross field by aligning it with one edge of a triangular patch and subsequently applies a mixed integer optimization to ensure consistency of the cross fields across adjacent patches. While this method effectively preserves local consistency of the cross field, it fails to ensure global consistency across the entire 3D model.
Power Fields~\cite{Power_Fields2013} and PolyVectors~\cite{PolyVectors_Diamanti2014} introduce convex smoothness energies for generating smooth N-RoSy fields. Power Fields~\cite{Power_Fields2013} achieves global optimality through a convex formulation, but relies on a nonlinear transformation that can introduce additional singularities and geometric distortion.
Instant Meshes (IM)~\cite{Instant_Meshes2015} and its variants~\cite{QuadriFlow2018, quadwild2021} employ extrinsic energy to guide cross field optimization, enabling the generation of visually high-quality fields. However, their strong reliance on local information often undermines global consistency and leads to undesirable singularities.

Most existing methods prioritize the smoothness of cross fields but fail to accurately align with curvature directions. Ideally, a high-quality cross field should balance smoothness with faithful alignment to principal curvature directions or sharp geometric features.
To this end, NeurCross~\cite{NeurCross2024} jointly optimizes the cross field and a neural signed distance field (SDF), where the zero-level set of the SDF serves as a proxy for the input surface. NeurCross~\cite{NeurCross2024} outperforms existing methods in terms of robustness to surface noise and geometric fluctuations, and alignment with curvature directions and sharp feature curves. However, its optimization process is computationally intensive, limiting its applicability in real-time or interactive scenarios.
Moreover, these methods rely on per-shape optimization, which is often time-consuming.
In contrast, our approach is data-driven and enables direct inference on unseen shapes without requiring additional optimization.

\paragraph{Data-driven Methods} 
As a pioneering effort in utilizing deep learning for generalized generation, \citet{DL2quadMesh2021} introduces a fully automated, learning-based approach for predicting direction fields. 
By combining the information from a global network encoding the global semantics, a local network encoding the local geometry information, and a set of local reference frames indicating the local triangle position and rotation, it can infer a frame field in a feed-forward manner.
However, this method relies on a global network (i.e., PointNet~\cite{QSMG17}) to encode the global domain knowledge of a single category (i.e., human-like body priors of quad meshes). Therefore it remains open as how to extend this method to shapes of different categories and even unseen categories.
This method also assumes shapes are aligned in a canonical space, thus even small non-rigid changes, such as head rotations in human models, may disrupt the predictions.

The more recent work Point2Quad~\cite{li2025point2quad} proposes a learning-based approach to extract quad meshes from point clouds by generating quad candidates via $k$-NN grouping and filtering them using an MLP-based classifier. While promising, this candidate-driven approach requires post-processing heuristics and may produce non-watertight or non-manifold meshes. Furthermore, this method tends to produce extra undesired singularities, as shown by our test. In contrast, our method employs a sparse, locality-aware encoder that captures intrinsic geometric features without relying on global semantics. This design enables robust generalization across diverse shape categories and resilience to non-rigid deformations or pose variations, without requiring canonical alignment.

\begin{figure*}[!t]
  \centering
   \begin{overpic}[width=\linewidth]{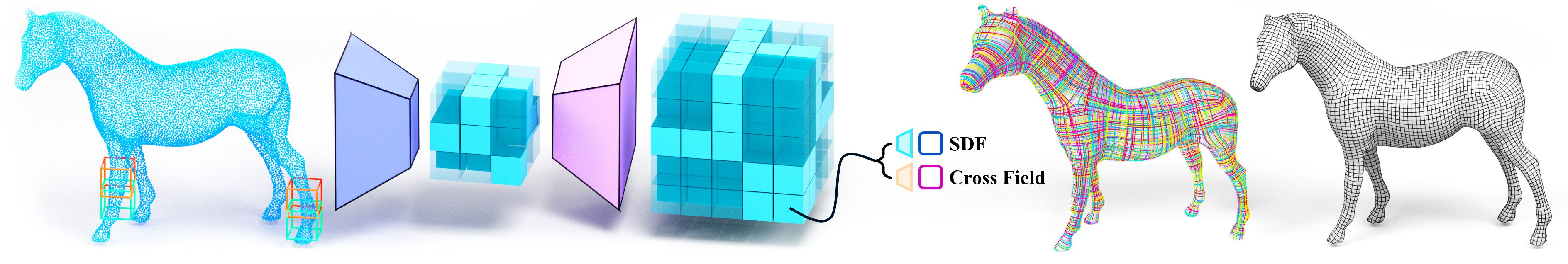}
    \put(8,-1){\small (a) Input}

    \put(23, 7.2){\large $\mathcal{E}$}
    \put(36.5, 7.2){\large $\mathcal{D}$}
    
    \put(25,-1){\small (b) Latent code $\mathcal{F}$}
    \put(40,-1){\small (c) High-res. feature}
    \put(55,-1){\small (d) Geo decoding}
    \put(69,-1){\small (e) Cross field}
    \put(87,-1){\small (f) Quad mesh}
    \end{overpic}
   \caption{Our network pipeline for learning and generating SDF and cross fields. Given a geometry input (mesh/point cloud) (a), we first encode the local geometry into sparse grids of latent embeddings (b), shown in light blue, and then decode these latent embeddings into a high-resolution feature grid (c), which can be interpreted (d) into SDF and cross fields (e) for downstream quad mesh generation (f).
   }
   \label{fig:method_overview}
\end{figure*}
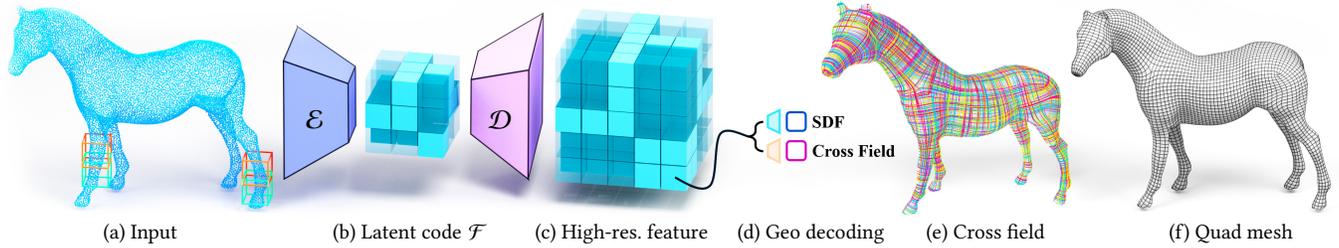

\subsection{Cross Field Applications}
Cross fields play a critical role in various geometry processing tasks. In quad mesh generation~\cite{BCE13_quad}, they guide parameterization to ensure quad edges align with principal curvature directions, producing meshes that better preserve surface geometry and curvature, enhancing both aesthetic and functional performance. For surface subdivision~\cite{subdivision2014}, cross fields are used to construct the base mesh topology by decomposing the parameter space, automatically determining singular points, and ensuring tangential continuity across boundaries. Similarly, in T-spline design~\cite{TSpline2017}, cross fields support the computation of global conformal parameterizations, from which aligned T-mesh domains are derived. These T-meshes enable the construction of smooth, piecewise rational surfaces for precise geometric modeling. In architectural geometry~\cite{architecturalGeometry2012}, cross fields are employed to create specialized planar quad meshes, including conic meshes, that achieve structural stability and moment-free equilibrium in steel-glass assemblies. By aligning mesh edges with curvature fields, these designs ensure self-supporting frameworks with minimal forces on the glass, supporting efficient and elegant constructions. Together, these applications highlight the versatility and importance of cross fields in geometry processing and design. In this work, we focus specifically on quadrilateral meshing, which serves as a foundational component for many of these applications.

\section{Method} 
Our goal is to efficiently generate a high-quality cross field of a given input point-cloud surface.
We propose \emph{CrossGen} based on an autoencoder architecture with novel designs of both encoder and decoder to enhance the generalization to unseen shapes. Fig.~\ref{fig:method_overview} shows the pipeline of \emph{CrossGen}. In the following, we describe the network architecture, loss functions, and training dataset. 

\begin{figure}[t]
  \centering
   \begin{overpic}[width=\linewidth]{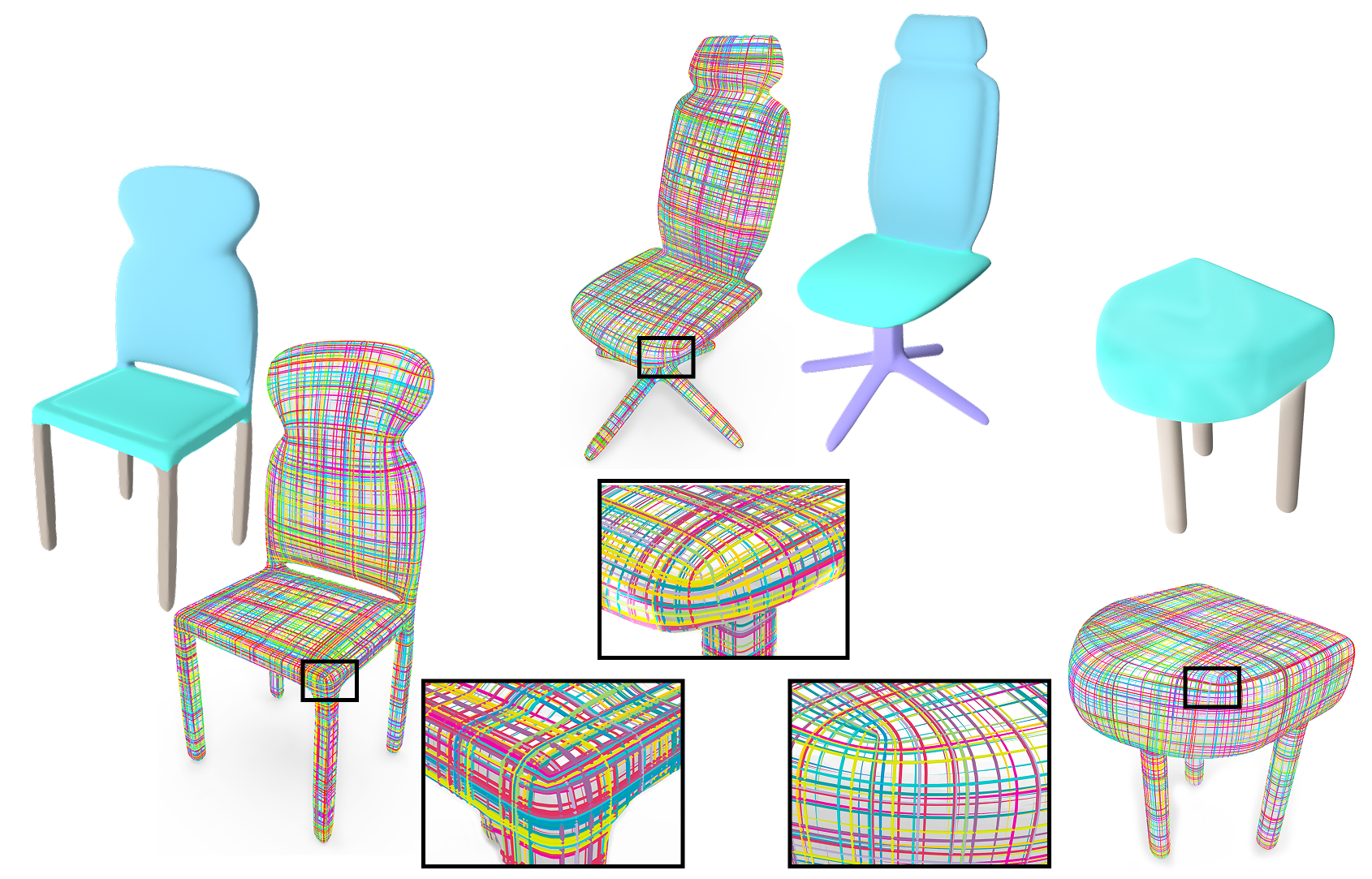}
   \end{overpic}
   \caption{
      Local geometric similarities across different shapes and their corresponding cross fields. Matching colors highlight geometrically similar regions. While the global structures differ significantly, the local geometries (highlighted windows) across the three shapes exhibit strong similarity. By leveraging a local encoder to capture these similar local patterns, our model generalizes more effectively across diverse shapes. 
    }
   \label{fig:encoder_explain}
\end{figure}

\subsection{Network Architecture}\label{sec:vae_cf}

\paragraph{Network Design}
\emph{CrossGen} adopts an autoencoder network architecture that takes a point-cloud surface as input and outputs the SDF of the input shape and a cross field on it. 
In order to ensure generalization across diverse shapes, 
we propose to use a shallow encoder based on a convolutional neural network (CNN) that focuses on local shape information without the need for learning global shape semantics. We then adopt a deep CNN decoder that takes into consideration of global geometry context for smooth and globally consistent generation of both SDFs and cross fields. We will shortly explain how to achieve these capabilities of learning local or global information using different ranges of the receptive field of a CNN.

Our design of using a local encoder and a global decoder is crucial for network generalization. 
On the one hand, shapes across different categories often appear globally different but share local similarities.
For example, different local parts such as the chairs and the table may share similar local geometry, as illustrated in Fig.~\ref{fig:encoder_explain}.
Using a local geometry encoding mechanism helps to capture these local similar patterns, improving generalization across these diverse categories as demonstrated in~\cite{mittal2022autosdf,rombach2022stablediffusion}.
On the other hand, geometry decoding necessitates the integration of global geometric context to ensure smoothness and consistency in predictions, thereby enabling the learning of a more generalized representation.

\paragraph{Encoder}
We adopt point clouds as the input surface format, which conveniently accommodates raw point-cloud surface as well as other surface representations, since any other surface representation can be converted to a point-cloud representation with a simple uniform sampling process.
The resulting input to the encoder~$\mathcal{E}$ is a 3D point set normalized to the unit cube.
Let the input point set be denoted by $ \mathcal{P}= \{ \boldsymbol{p} \mid \boldsymbol{p} \in \mathbb{R}^3 \}$, where each point~$\boldsymbol{p}$ is assumed to be associated with its corresponding surface unit normal vector~$ \boldsymbol{n}_{\boldsymbol{p}} \in \mathbb{R}^3$ as an input feature. These surface normal vectors can be extracted from the original continuous surface if provided as input, or estimated using \citet{normal_Xu506} for a raw point-cloud surface.
The point set~$\mathcal{P}$ is first quantized onto the vertices of a high-resolution cubic grid~$\Omega$, with each point~$\boldsymbol{p}$ mapped to its nearest voxel vertex. Voxels containing at least one quantized point are designated as active, whereas all others are regarded as empty and omitted from subsequent sparse convolution operations. To preserve fine geometric details, a sufficiently high grid resolution~$\Omega$ is desirable. However, increasing the resolution inevitably incurs substantial memory and computational overhead. In practice, we set $\Omega=256^3$, which offers a balanced compromise between fidelity and efficiency.
This quantized data is subsequently encoded into 16-dimensional latent embeddings by a linear projection layer and fed into the encoder~$\mathcal{E}$, which uses a sparse convolutional neural network (CNN) architecture~\cite{choy2019sparsecnn} -- a crucial choice for efficiently processing high-resolution 3D data while significantly reducing computational overhead.

The encoder consists of a sequence of convolutional layers and residual blocks that progressively downsample the input representation. As data propagates through the network, with the successive spatial resolutions of the feature maps being $256^3$, $128^3$, $64^3$, $32^3$, and $16^3$, and corresponding feature dimensions being 16, 32, 64, 128, and 128. 
Each downsampling step uses a single convolutional layer to ensure a small receptive field and encourage the encoding of local geometric structures rather than global semantic features.
At the network bottleneck, i.e. latent space, the 3D shape is compactly represented as a set of latent patch descriptors~$\mathcal{F} = \{ \boldsymbol{f} \mid \boldsymbol{f} \in  \mathbb{R}^m \}$, where $m$ ($m=128$ by default) denotes the feature dimensionality, as shown in Fig.~\ref{fig:method_overview}~(b). 

\paragraph{Decoder}
The decoder~$\mathcal{D}$, which also utilizes the sparse CNN architecture~\cite{choy2019sparsecnn}, comprises a sequence of convolutional upsampling layers and residual blocks, progressively increasing the spatial resolution of the feature maps to $16^3$, $32^3$, and $64^3$, with the corresponding feature dimensions decreasing from 128 to 64 and then to 32.
This process yields a high-resolution feature grid, as shown in Fig.~\ref{fig:method_overview}~(c).
At each resolution level, multiple convolutional layers are employed to expand the receptive field, enabling the propagation of geometric information to neighboring regions and enhancing feature aggregation for accurate SDF and cross field decoding.

\begin{wrapfigure}{r}{4cm}
\vspace{-5mm}
  \hspace*{-4mm}
  \centerline{
  \begin{overpic}[width=40mm]{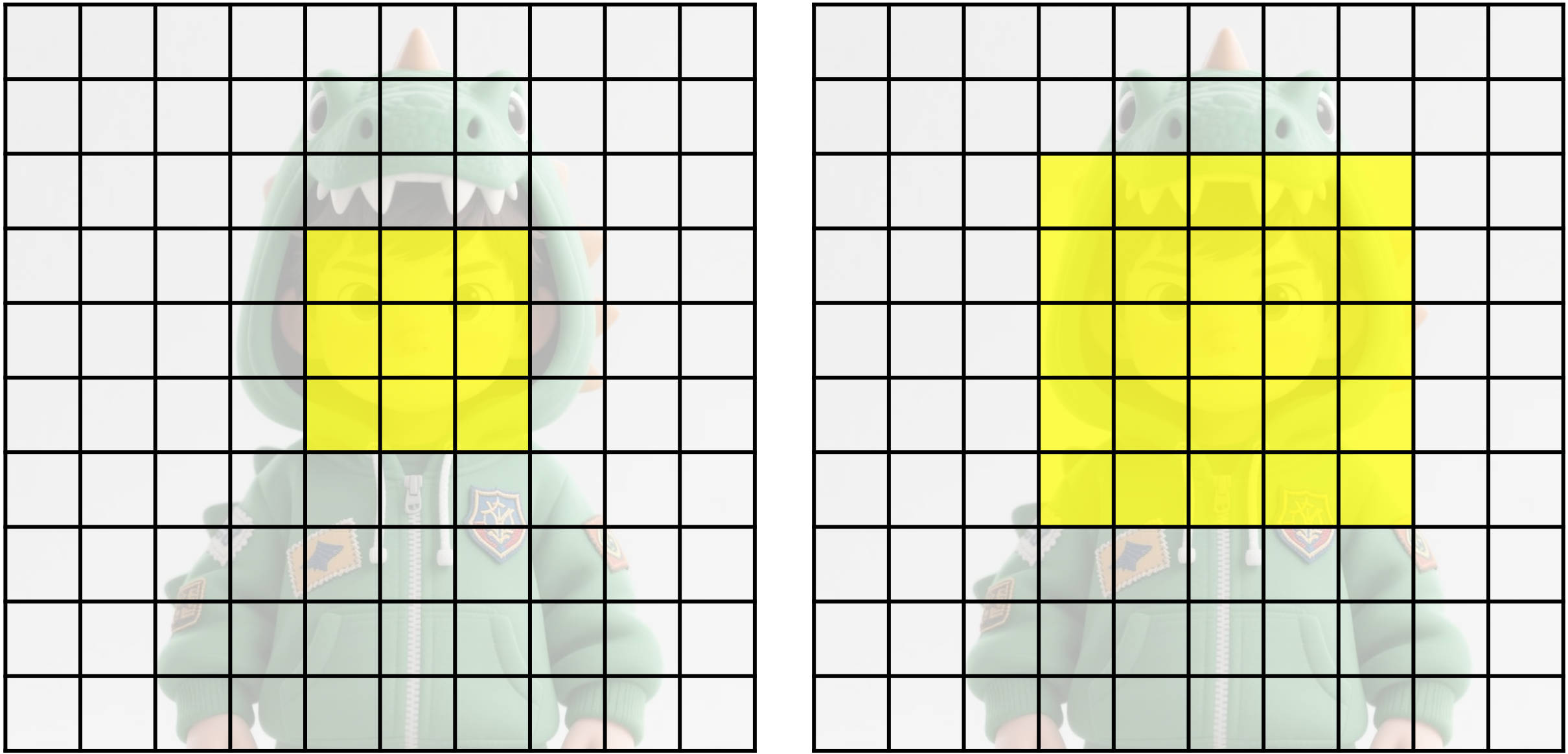}
        \put(13,-10){\small{Small RF}}
        \put(65,-10){\small{Large RF}}
  \end{overpic}
  }
  \vspace{1mm}
\end{wrapfigure}
\paragraph{Receptive Field}
The receptive field~(RF) of a CNN describes how much context of the input a network can ``see'' when making predictions at a given location. It defines the geometric context available for encoding the input point cloud surface and predicting the output, i.e. SDFs or cross fields. 
As shown in the inset for 2D illustration, smaller receptive fields focus on local details, while larger ones capture broader shape context.
As discussed in Section~\ref{sec:vae_cf}, although shapes from different categories may vary globally, they often share local geometric similarities. To leverage this, we employ a small receptive field in the encoder to focus on local geometry (see Fig.~\ref{fig:encoder_explain}). Meanwhile, we adopt a large receptive field in the decoder to aggregate broader context, which promotes smoothness and global coherence in the predicted cross fields.
In other words, locality is governed by the receptive field of each voxel in the sparse CNN, rather than by explicit patching or segmentation. The encoder’s use of small kernels and shallow layers ensures that voxel features retain fine-grained geometric detail. Moreover, because sparse convolutions operate over a global voxel grid, receptive fields of adjacent voxels naturally overlap, promoting smooth regional transitions. This convolutional hierarchy implicitly aggregates neighboring features, eliminating the need for explicit fusion operations such as averaging.

\paragraph{SDF and Cross Field Modules}
Subsequent to the construction of our high-resolution volumetric feature representation (see Fig.~\ref{fig:method_overview}~(c)), we employ trilinear interpolation to extract two 32-dimensional feature vectors: one at the input points~$\mathcal{P}= \{ \boldsymbol{p} \mid \boldsymbol{p} \in \mathbb{R}^3 \}$ for cross field prediction, and another at query points~$\mathcal{Q}= \{ \boldsymbol{q} \mid \boldsymbol{q} \in \mathbb{R}^3 \}$ for SDF prediction. 
The query set~$\mathcal{Q}$ is sampled from a thin shell surrounding the zero-level set of the SDF, composed of points whose ground truth SDF magnitudes 
fall below a prescribed threshold~$\epsilon$ (we use $\epsilon=0.02$ by default). Note that~$\mathcal{P}$ constitutes a proper subset of~$\mathcal{Q}$.

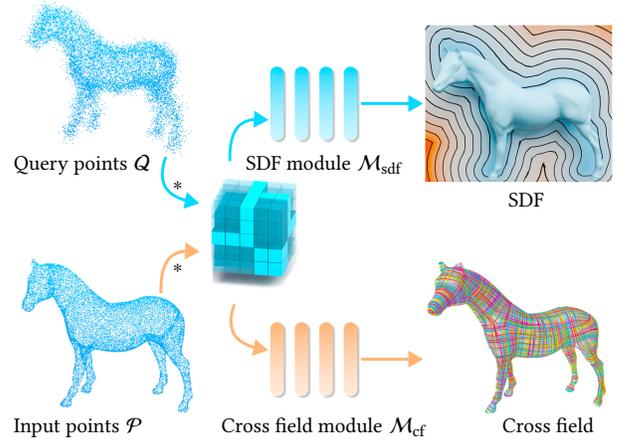
\begin{figure}[t]
  \centering
   \begin{overpic}[width=0.95\linewidth]{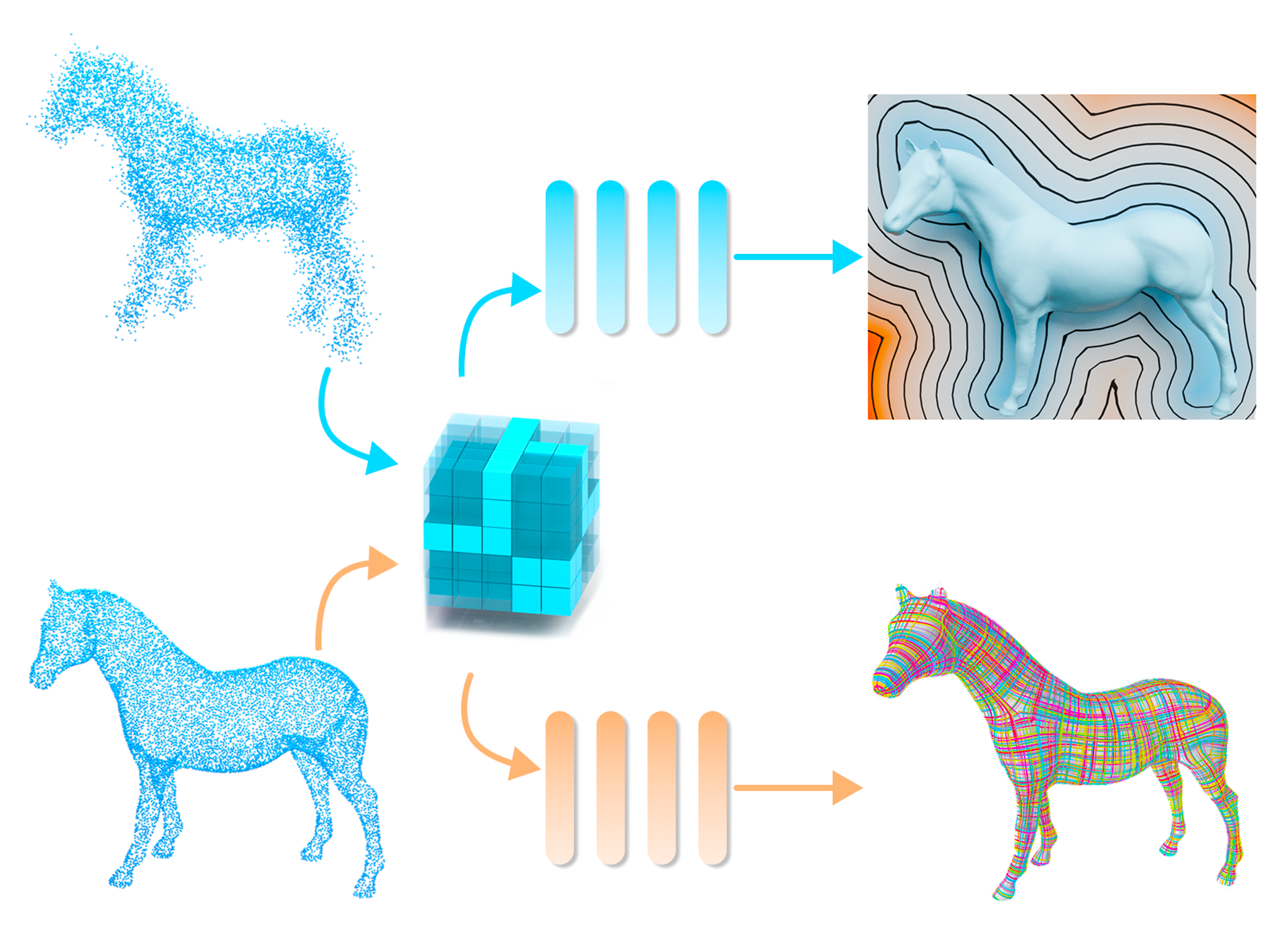}
    \put(1,44){\small Query points~$\mathcal{Q}$}
    \put(1,1){\small Input points~$\mathcal{P}$}
    \put(39,44){\small SDF module~$\mathcal{M}_\text{sdf}$}
    \put(35,1){\small Cross field module~$\mathcal{M}_\text{cf}$}
    \put(81,1){\small Cross field}
    \put(82,38){\small SDF}
    \put(27,40.5){\small $\mathcal{*}$}
    \put(27,27){\small $\mathcal{*}$}
   \end{overpic}
   \caption{
   Overview of our SDF and cross field modules. Both modules are implemented as MLPs that decode the high-resolution feature grid into SDFs and cross field, respectively. $\mathcal{*}$ represents trilinear interpolation operation.
   }
   \label{fig:mlp_modules}
\end{figure}

These interpolated features are fed into two distinct MLP decoders~(see Fig.~\ref{fig:mlp_modules}).
One MLP decoder,~$\mathcal{M}_\text{cf}(\boldsymbol{p}) \in \mathbb{R}^3$, maps the feature vector at point~$\boldsymbol{p}$ to one cross field direction, while another,~$\mathcal{M}_\text{sdf}(\boldsymbol{q}) \in \mathbb{R}$, maps the feature vector at point~$\boldsymbol{q}$ to a SDF value.
In essence,~$\mathcal{M}_\text{cf}$ decodes the interpolated feature vector at input points~$\mathcal{P}$ into directional cross field vectors, while~$\mathcal{M}_\text{sdf}$ interprets features in the thin shell~$\mathcal{Q}$ into SDF values.

\subsection{Loss Function}
For training of \emph{CrossGen}, we draw inspiration from previous works \cite{StructRe2025, NeurCross2024, rombach2022stablediffusion} and design a composite loss function comprising four terms:
\begin{enumerate}
    \item {\em Occupancy Loss}: Denoted  $\mathcal{L}_\text{o}$, a term for multi-resolution volumetric fidelity that is applied to sparse grid representations across the successive layers of the decoder.
    \item {\em Cross Field Loss}: Denoted $\mathcal{L}_\text{cf}$, a term for enforcing cross field consistency at the input point cloud~$\mathcal{P}$.
    \item {\em SDF Loss}: Denoted $\mathcal{L}_\text{sdf}$, a term for predicting SDF values at queried points~$\mathcal{Q}$ within the thin-shell space around the zero-level set of the SDF.
    \item {\em Latent Space Regularization}: Denoted  $\mathcal{L}_\text{kl}$, a term that enforces structural coherence in the latent representation space for optimized feature encoding.
\end{enumerate}

These individual loss terms and their corresponding weights are explained in detail below.

\paragraph{Occupancy Loss}
Since raw point clouds are often noisy, sparse, and incomplete, direct occupancy estimation can be unreliable; therefore, we incorporate an occupancy loss into our framework. This loss also acts as a regularizer during training, encouraging the network to learn stronger geometric priors across shapes and thereby maintain local geometric consistency.

In our implementation, both the encoder~$\mathcal{E}$ and decoder~$\mathcal{D}$ adopt the sparse CNN architecture~\cite{choy2019sparsecnn} for good computational and memory efficiency. 
In the encoder~$\mathcal{E}$, empty voxels are directly removed during the downsampling process, keeping only the occupied regions for efficient feature extraction.
During upsampling by the decoder~$\mathcal{D}$, newly generated fine-grained sub-voxels, denoted as~$\mathcal{V}$, are retained and marked as occupied only if they intersect with the thin shell region surrounding the zero-level set of the SDF; the rest are discarded.
To supervise voxel occupancy prediction at each resolution level of the output of the decoder~$\mathcal{D}$, we introduce an occupancy loss~$\mathcal{L}_\text{o}$, defined as the discrepancy between the predicted occupancy values~$\mathcal{D}_\text{o}(\boldsymbol{v})$ and the ground truth occupancy labels $\mathcal{G}_\text{o}(\boldsymbol{v})$.
\begin{equation}
	\mathcal{L}_\text{o} = -\frac{1}{|\mathcal{V}|}\sum_{\boldsymbol{v} \in \mathcal{V}}{\textrm{BCE}\bigl(\mathcal{D}_\text{o}(\boldsymbol{v}), \mathcal{G}_\text{o}(\boldsymbol{v})\bigr)},
\end{equation}
where \textrm{BCE}(·) is the binary cross entropy, that is:
\begin{equation}
	\textrm{BCE}(\boldsymbol{x}, \boldsymbol{y}) = \boldsymbol{y} \log(\boldsymbol{x}) + (1-\boldsymbol{y}) \log(1-\boldsymbol{x}).
\end{equation}

\paragraph{Cross Field Loss}
\begin{wrapfigure}{r}{4cm}
\vspace{-1mm}
  \hspace*{-4mm}
  \centerline{
  \begin{overpic}[width=40mm]{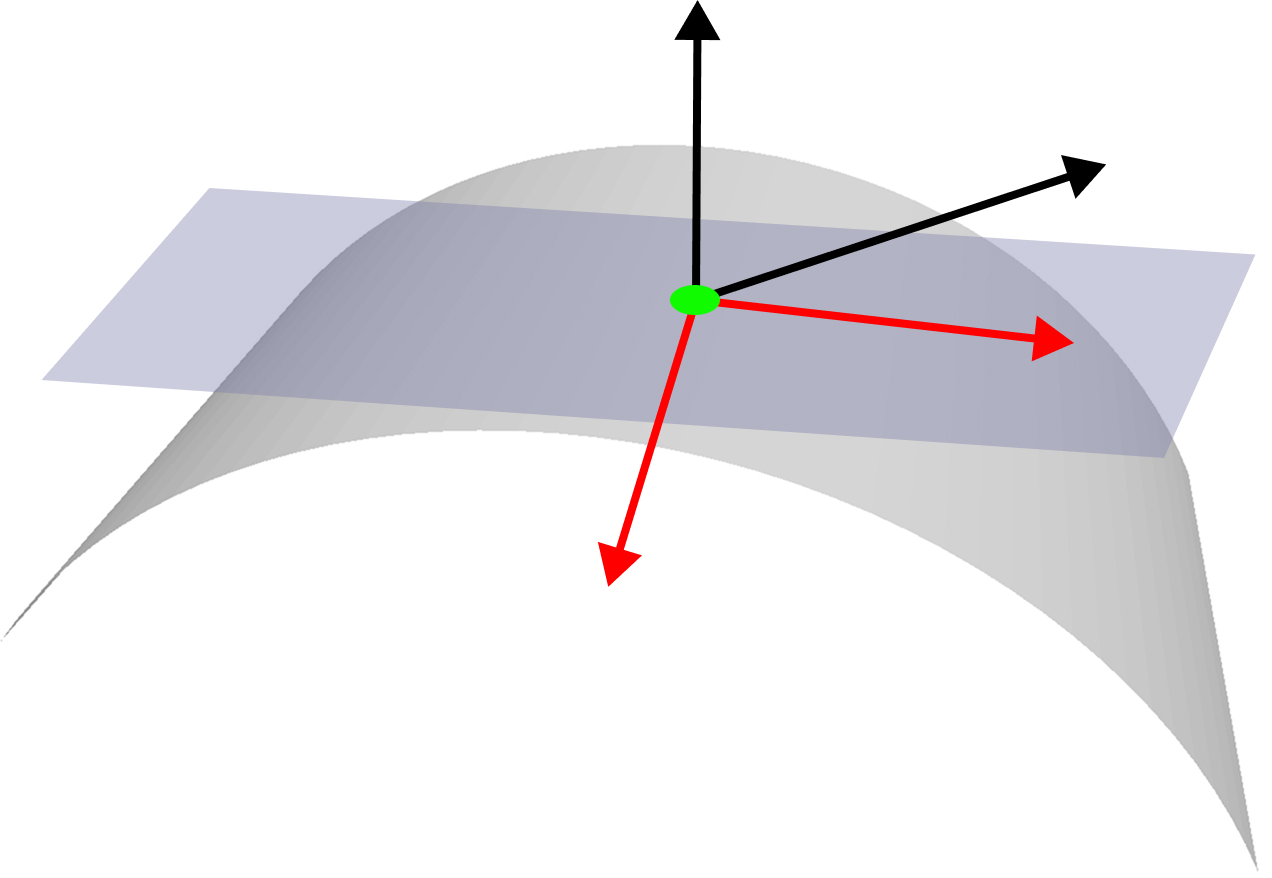}
        \put(55,38){$\boldsymbol{p}$}
        \put(70,62){$\mathcal{M}_\text{cf}(\boldsymbol{p})$}
        \put(81.5,34){$\bar{\boldsymbol{\alpha}}_{\boldsymbol{p}}$}
        \put(51,20){$\boldsymbol{\beta}_{\boldsymbol{p}}$}
        \put(44.5,69){$\boldsymbol{n}_{\boldsymbol{p}}$}
        \put(-3,46){\small{tangent plane}}
  \end{overpic}
  }
  \vspace*{-1mm}
\end{wrapfigure}
Given a surface point~$\boldsymbol{p}$, the cross field branch $\mathcal{M}_\text{cf}$ predicts one direction vector of the cross field at~$\boldsymbol{p}$, denoted by~$\mathcal{M}_\text{cf}(\boldsymbol{p})$.
This predicted vector~$\mathcal{M}_\text{cf}(\boldsymbol{p})$ needs to be constrained to lie within the tangent plane at~$\boldsymbol{p}$. Hence, we  project~$\mathcal{M}_\text{cf}(\boldsymbol{p})$ onto the tangent plane defined by~$\boldsymbol{p}$ and its corresponding GT unit normal vector~$\boldsymbol{n}_{\boldsymbol{p}}$.
As illustrated in the inset figure, we consider a local region centered at~$\boldsymbol{p}$.
Using the unit normal~$\boldsymbol{n}_{\boldsymbol{p}}$, we project the predicted vector onto the tangent plane, resulting in~$\bar{\boldsymbol{\alpha}}_{\boldsymbol{p}}$, a direction vector that lies within the tangent plane at~$\boldsymbol{p}$. 
That is
\begin{equation}
	\bar{\boldsymbol{\alpha}}_{\boldsymbol{p}} = \mathcal{M}_\text{cf}(\boldsymbol{p})-\left(\mathcal{M}_\text{cf}(\boldsymbol{p}) \cdot \boldsymbol{n}_{\boldsymbol{p}}\right)\boldsymbol{n}_{\boldsymbol{p}}=(\boldsymbol{I} - \boldsymbol{n}_{\boldsymbol{p}}^\text{T}\boldsymbol{n}_{\boldsymbol{p}})\mathcal{M}_\text{cf}(\boldsymbol{p}).
\end{equation}
We then normalize vector~$\bar{\boldsymbol{\alpha}}_{\boldsymbol{p}}$ to obtain the unit vector~$\boldsymbol{\alpha}_{\boldsymbol{p}}$, which represents one cross field direction at point~$\boldsymbol{p}$ predicted by our network.
The second direction~$\boldsymbol{\beta}_{\boldsymbol{p}}$ is  
computed as $\boldsymbol{\beta}_{\boldsymbol{p}} = \boldsymbol{\alpha}_{\boldsymbol{p}} \times \boldsymbol{n}_{\boldsymbol{p}}$.
Thus, the cross field at point~$\boldsymbol{p}$ predicted by our \emph{CrossGen} is given by the pair~$(\boldsymbol{\alpha}_{\boldsymbol{p}}, \boldsymbol{\beta}_{\boldsymbol{p}})$, both of which are unit vectors lying in the tangent plane.

During training, to learn the cross field~$(\boldsymbol{\alpha}_{\boldsymbol{p}}, \boldsymbol{\beta}_{\boldsymbol{p}})$, we provide supervision by comparing it against the ground truth cross field~$(\boldsymbol{\mu}_{\boldsymbol{p}}, \boldsymbol{\nu}_{\boldsymbol{p}})$. However, the exact correspondence between the two pairs is ambiguous due to their rotational symmetry. 
Following~\citet{NeurCross2024}, it suffices to require that one vector~(i.e. $\boldsymbol{\alpha}_{\boldsymbol{p}}$) from ~$(\boldsymbol{\alpha}_{\boldsymbol{p}}, \boldsymbol{\beta}_{\boldsymbol{p}})$ be aligned, being parallel or perpendicular, to both vectors in~$(\boldsymbol{\mu}_{\boldsymbol{p}}, \boldsymbol{\nu}_{\boldsymbol{p}})$. It can be proved that this requirement is fulfilled if and only if~$\left\| \boldsymbol{\alpha}_{\boldsymbol{p}} \cdot \boldsymbol{\mu}_{\boldsymbol{p}} \right\| + \left\| \boldsymbol{\alpha}_{\boldsymbol{p}} \cdot \boldsymbol{\nu}_{\boldsymbol{p}} \right\|$ attains its minimum value of 1~\cite{NeurCross2024}.
Hence, the cross field loss term can be written as:
\begin{equation}
\label{equ:loss_cf}
\mathcal{L}_\text{cf} = \frac{1}{|\mathcal{P}|} \sum_{\boldsymbol{p} \in \mathcal{P}} \left( \left\| \boldsymbol{\alpha}_{\boldsymbol{p}} \cdot \boldsymbol{\mu}_{\boldsymbol{p}} \right\| + \left\| \boldsymbol{\alpha}_{\boldsymbol{p}} \cdot \boldsymbol{\nu}_{\boldsymbol{p}} \right\| - 1 \right),
\end{equation}
where~$\left\| \cdot \right\|$ denotes the absolute value.

\paragraph{SDF Loss}
The SDF loss at point~$\boldsymbol{q} \in \mathcal{Q}$, measuring the discrepancy between the predicted value~$\mathcal{M}_{\text{sdf}}(\boldsymbol{q})$ and the ground truth SDF~$\mathcal{G}_\text{sdf}(\boldsymbol{q})$, is defined as:
\begin{equation}
	\mathcal{L}_\text{sdf} = \frac{1}{|\mathcal{Q}|}\sum_{\boldsymbol{q}\in \mathcal{Q}}{\left\| \mathcal{M}_\text{sdf}(\boldsymbol{q}) - \mathcal{G}_\text{sdf}(\boldsymbol{q}) \right\|}. 
\end{equation}

\paragraph{Latent Space Regularization}
To prevent high variance and stabilize the latent space~$\mathcal{F}=\{ \boldsymbol{f} \mid \boldsymbol{f} \in  \mathbb{R}^{128} \}$, we incorporate a KL divergence regularization term~$D_\text{KL}$~\cite{kingma2013vae}, following \cite{mo2019structurenet,rombach2022stablediffusion}.
This constrains the latent space distribution~$\mathcal{N}(\mathbb{E}(f), \sigma^2(f))$ toward a standard normal distribution~$\mathcal{N}(0, 1)$. 
Both empirical evidence and established theoretical results~\cite{van2017vqvae, yan2022shapeformer, mittal2022autosdf,kingma2013vae} demonstrate that this distributional constraint ensures a topologically coherent and smooth latent space \cite{mo2019structurenet,rombach2022stablediffusion}, substantially reducing information entropy during continuous shape parametrization transformations.
Hence, we define the latent space regularization as follows:
\begin{equation}
\mathcal{L}_\text{kl} = \frac{1}{|\mathcal{F}|} \sum_{\boldsymbol{f} \in \mathcal{F}} D_\text{KL}\left(\mathcal{N}(\mathbb{E}(f), \sigma^2(f)), \mathcal{N}(0, 1)\right),
\end{equation}
where~$\mathbb{E}(\cdot)$ denotes the mean and~$\sigma^2(\cdot)$ the variance.

\paragraph{Total Loss}
Finally, the overall training loss $\mathcal{L}$ is defined as the sum of the occupancy loss, cross field loss, SDF loss, and the latent space regularization term:
\begin{equation}
  \mathcal{L} = \lambda_\text{o} \cdot \mathcal{L}_\text{o} + \lambda_\text{cf} \cdot \mathcal{L}_\text{cf} + \lambda_\text{sdf} \cdot \mathcal{L}_\text{sdf} + \lambda_\text{kl} \cdot \mathcal{L}_\text{kl},
\label{eq:loss_patchlatent}
\end{equation}
where $\lambda_\text{o}=1, \lambda_\text{cf}=1, \lambda_\text{sdf}=1, \lambda_\text{kl}=1e^{-6}$ denote the corresponding weights of different loss terms to balance and stabilize the training process.

\subsection{Dataset for Training}\label{sec:data_prepare}
We have constructed a large-scale dataset of 3D shapes with ground-truth SDFs and cross fields for training our network and evaluating our method. To the best of our knowledge, there is no public dataset of 3D shapes that offers the ground truth of paired SDF and cross fields, which are needed for our training. For all the shapes in the dataset, their ground truth SDFs are computed using the Truncated Signed Distance Function~(TSDF)~\cite{TSDF}, and their cross fields are computed using NeurCross~\cite{NeurCross2024}.

To ensure geometric diversity for training a highly generalizable and robust network, the dataset contains 1,700 distinct shapes, including smooth organic surfaces, common object shapes~(e.g. chairs, airplanes), and CAD models from the following sources:
\begin{enumerate}
\item {\em Smooth Organic Shapes}:
These shapes are drawn from DeformingThings4D~\cite{li20214deform4d}, a synthetic dataset comprising 1,972 animation sequences across 31 categories of humanoid and animal models, with each category having 100 posed instances.
We randomly select 10 models per category, yielding a total of {\bf 310 models}.
\item {\em Common Object Shapes}: 
We sample 550 shapes from ShapeNetCore~\cite{ShapeNet}, which consists of 51,300 clean 3D models spanning 55 categories, by randomly selecting 10 shapes in each category, yielding {\bf 550 models}.
Additionally, we randomly select {\bf 530 models} from Thingi10K \cite{Thingi10K}, a dataset known for its diverse collection of geometrically intricate 3D shapes.
\item CAD Models: 
We select {\bf 310 models} from the ABC dataset \cite{Koch_2019_abc}, which comprises one million CAD models with explicitly parameterized curves and surfaces.
\end{enumerate}

\begin{figure}[t]
  \centering
   \begin{overpic}[width=\linewidth]{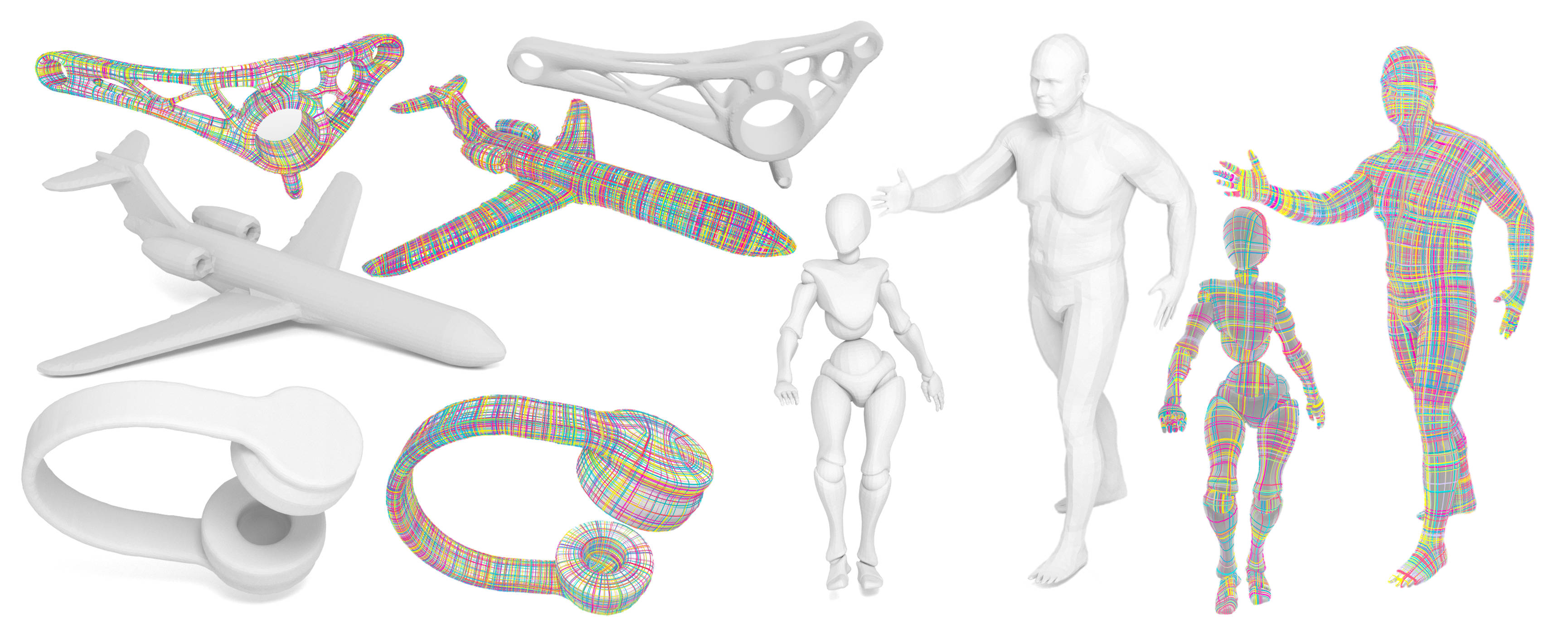}
   \end{overpic}
   \caption{
      Sample shapes from our dataset. Our dataset includes over 10,000 shapes with high-quality cross fields and SDFs. Both the cross fields and corresponding SDF surfaces are visualized in the figure.
    }
   \label{fig:dataset}
\end{figure}

To enhance model resilience and promote orientation-invariant cross field predictions, the {\bf 1,700 distinct shapes} in the dataset are augmented with random rotations, where each shape is rotated by a random angle about a random axis, yielding over 10,000 shapes. All the shapes are normalized to the cube~$[-0.5, 0.5]^3$ to ensure consistent scaling across the shapes.
Fig.~\ref{fig:dataset} presents examples of cross field and SDF surfaces included in our dataset.
The dataset is partitioned at the distinct shape level using a 9:1 train–test split.

\subsection{Application to Quad Meshing}
\label{sec:local_encoder}

\paragraph{Quad Mesh Extraction}
We now discuss how to use the cross field predicted by \emph{CrossGen} to produce a quad mesh of the given shape. 
Since the input to \emph{CrossGen} is a discrete point-cloud surface but existing quad mesh extraction methods require a mesh surface to operate on, we use the SDF predicted by \emph{CrossGen} to provide such a continuous surface for quad meshing. Specially, we obtain a triangle mesh surface of the zero-level set of the predicted SDF using the Marching Cubes method~\cite{MC}. Then we compute the center points of the triangle faces and query their corresponding features from the high-resolution grid (Fig.~\ref{fig:method_overview} (c)). These features are then passed through the cross field branch to predict the cross fields. Finally, the predicted cross field and reconstructed surface mesh are used together to extract a field-aligned quad mesh.
If the input shape is provided as a triangle mesh, we directly perform quad mesh extraction using the predicted cross field, without needing to reconstruct the surface through the SDF branch.

For quad mesh extraction, we implement a two-step pipeline similar to~\citet{NeurCross2024, MIQ2009, DL2quadMesh2021}. We first employ the global seamless parameterization algorithm from the libigl~\cite{libigl2017}, which establishes parametric alignment with the predicted cross field. Then we utilize the libQEx~\cite{libQEX13} to extract the quadrilateral tessellation from the parameterized representation.

\paragraph{Novel Quad Mesh Generation}
With our unified geometry representation, 3D shapes are encoded into a sparse latent space that captures both geometric structure and cross field information. This latent space  provides a natural foundation for generative modeling~\cite{zheng2023LASDiffusion}. By incorporating a diffusion model within this latent, \emph{CrossGen} can be extended to synthesize novel shapes {\em and} compute their cross fields in an end-to-end manner, providing an efficient way of quad meshing of the novel shapes.  We provide further details and evaluations of this generative capability of \emph{CrossGen} in Sec.~\ref{sec:Generated_Shapes}.

\section{Experiments}

\paragraph{Implementation Details} 
\label{sec:metrics}
We utilize distinct evaluation metrics to measure the quality of the generated cross field and the resulting quad mesh.
(1)
To evaluate the quality of the generated cross field, we employ two metrics: angular error~(AE) relative to the principal curvature directions and computational time.
The definition of AE is similar to that of Eq.~\ref{equ:loss_cf}, where $AE \in [0, \sqrt{2} - 1]$. 
(2)
To assess the quality of the generated quad meshes, we adopt five standard metrics following~\citet{QuadriFlow2018} and ~\citet{NeurCross2024}: area distortion (Area), angle distortion (Angle), number of singularities (\# of Sings), Chamfer Distance (CD), and Jacobian Ratio (JR).
Area distortion (Area), scaled by~$10^4$, measures the standard deviation of quadrilateral face areas, reflecting the uniformity of element sizing.
Angle distortion (Angle), is computed as $\sqrt{\frac{1}{N}\sum_i(\phi_i - \frac{\pi}{2})^2}$, where $\phi_i$ denotes interior angles and~$N$ is the total number of angles; it captures deviations from the ideal right angles.
Chamfer Distance (CD), scaled by~$10^4$ and measured using the $L_1$ norm, evaluates geometric similarity between the predicted and ground truth surfaces.
Jacobian Ratio (JR) quantifies local deformation uniformity by computing the ratio between the smallest and largest Jacobian determinants across element corners, ranging from 0 (degenerate) to 1 (perfect parallelogram).
The number of singularities counts irregular vertices in the quad mesh, reflecting topological regularity.

During training, we uniformly sample 150,000 points from the ground-truth mesh as input and apply random point dropping as an augmentation strategy. We trained our model on a system equipped with 8 NVIDIA GeForce RTX 4090 GPUs, each with 24 GB of memory. The training was performed for 2,000 epochs using the Adam optimizer~\cite{adam2014} with a learning rate of $1 \times 10^{-4}$ and a batch size of 16. The full training process took approximately 15 days to converge on the complete dataset.

\begin{table}[t]
    \centering
    \caption{Quantitative comparison of cross fields with baseline methods. The ``all'' refers to the entire test set, where shapes average 10,000 vertices and 20,000 faces. The ``high-res.'' denotes a subset of 50 test shapes, each with over 100,000 faces. The ``Angular Error (AE)'' quantifies the deviation between the principal curvature directions and the generated cross field across different methods. In the time metric, ``s'' denotes seconds.
    Within each column, the best scores are highlighted with bold and underline (\under{best}).
    }
    \resizebox{0.98\linewidth}{!}{
    \begin{tabular}{l|ccc}
    \hline\noalign{\smallskip}
    \multicolumn{1}{c|}{} & \multirow{2}{*}{ Angular Error (AE)~$\downarrow$} & \multicolumn{2}{c}{Time (s)~$\downarrow$} \\
    &   & all   & high-res. \\ 
    \noalign{\smallskip}
    \hline\noalign{\smallskip}
    MIQ~\footnotesize{~\cite{MIQ2009}} & 0.17 & 265.68 & 1201.38 \\
    Power Fields~\footnotesize{~\cite{Power_Fields2013}} &  0.11 &3.13 & 6.31 \\
    PolyVectors~\footnotesize{~\cite{PolyVectors_Diamanti2014}} & 0.10& 1.41 & 2.17 \\
    IM~~\footnotesize{\cite{Instant_Meshes2015}} & 0.07 & 2.13 & 5.28\\
    QuadriFlow~~\footnotesize{\cite{QuadriFlow2018}} & 0.07 & 2.37 &  5.58\\
    NeurCross~~\footnotesize{\cite{NeurCross2024}} & \under{0.05} & 360.58 & 912.46 \\
    \hline
    \textbf{CrossGen (Ours)} & \under{0.05} &  \under{0.067} & \under{0.079} \\
    \hline
    \end{tabular}
    }
    \label{tab:comp_cf}
\end{table}

\begin{figure*}[t]
  \centering
   \begin{overpic}[width=0.98\linewidth]{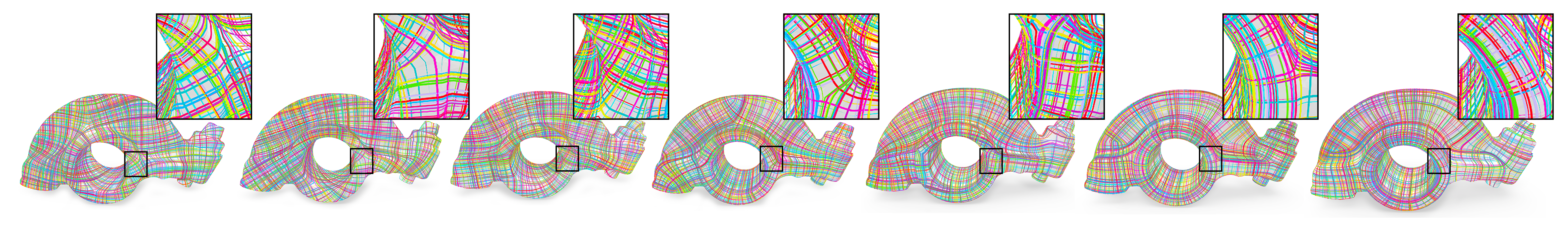}
        \put(5,-1.5){\small MIQ}
        \put(17,-1.5){\small Power Fields}
        \put(30,-1.5){\small PolyVectors}
        \put(46,-1.5){\small IM}
        \put(58,-1.5){\small QuadriFlow}
        \put(73,-1.5){\small NeurCross}
        \put(85,-1.5){\small CrossGen (Ours)}
   \end{overpic}
   \caption{
   Cross fields generated by six state-of-the-art methods and our CrossGen. Both CrossGen and NeurCross~\cite{NeurCross2024}, the ground truth provider for our dataset, produce smooth cross fields that align well with the principal curvature directions.
   }
   \label{fig:compare_cf}
\end{figure*}
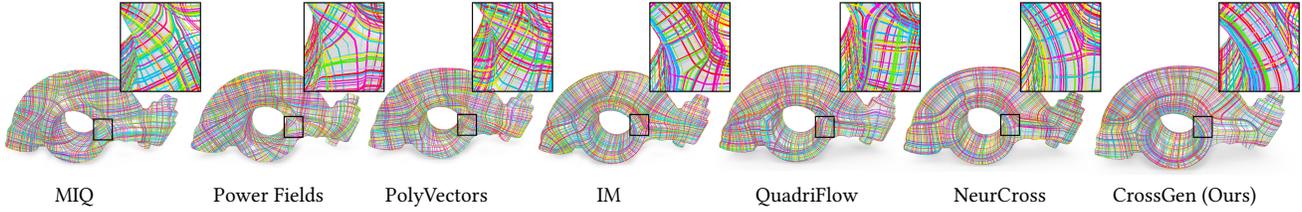

\subsection{Comparisons}\label{sec:quad_infer}
Quantitatively comparing our method is difficult because there are no publicly available learning-based methods for cross field generation in a feed-forward manner. 
However, to provide a comprehensive understanding of \emph{CrossGen}'s performance, we compare it with several state-of-the-art optimization-based methods, including  MIQ~\cite{MIQ2009}, Power Fields~\cite{Power_Fields2013}, PolyVectors~\cite{PolyVectors_Diamanti2014}, IM~\cite{Instant_Meshes2015}, QuadriFlow~\cite{QuadriFlow2018}, and NeurCross~\cite{NeurCross2024}.
For all the methods, except MIQ~\cite{MIQ2009}, we use their official open-source implementations with default parameters.
As MIQ~\cite{MIQ2009} does not provide source code, we adopt its implementation from libigl~\cite{libigl2017}.
Since all the baselines take mesh surfaces input, we also test the option of using the input mesh at the stage of quad mesh extraction. 

\paragraph{Cross Field Evaluation}   
Evaluating cross fields is inherently challenging, as principal curvature directions may become ambiguous in noisy or flat regions. To ensure rigorous and reliable assessment, we therefore select 50 smooth, noise-free shapes with well-defined principal directions, which minimizes ambiguity and provides a trustworthy reference for evaluation.
It should be noted that these
50 smooth, noise-free models are used exclusively for computing the angular error reported in Tab.~\ref{tab:comp_cf}, which
presents a quantitative comparison between our method and the baseline approaches. 
On the test data, our method and NeurCross~\cite{NeurCross2024} achieve the lowest angular error (AE) compared to the principal curvature directions, outperforming all other baselines.
Benefiting from the direct inference design, \emph{CrossGen} achieves the fastest average runtime, outperforming NeurCross~\cite{NeurCross2024} by approximately 5382 times on the full test data.

Note that only NeurCross~\cite{NeurCross2024} and our method are GPU-accelerated; all other baselines run on the CPU. To further evaluate performance under high-resolution inputs, we tested 50 randomly selected shapes from the test set, each with over 100,000 faces.
These 50 models are different from the 50 smooth, noise-free models used to compute the angular error. 
As shown in Tab.~\ref{tab:comp_cf}, our method maintains a consistent runtime regardless of input resolution.
In contrast, all CPU-based methods show significantly increased runtime with higher resolution, while NeurCross~\cite{NeurCross2024}, despite running on the GPU, is over 11,550 times slower than \emph{CrossGen} and becomes increasingly constrained by computational overhead at higher resolutions.
Fig.~\ref{fig:compare_cf} presents a visual comparison of the cross fields generated by various methods.

\begin{table}
    \centering
    \caption{
    Quantitative comparisons of quad meshes with baseline methods. All methods produced quad meshes with an average of approximately 10,000 vertices and 20,000 faces. Note that NeurCross is the method used to generate our training data.
    Within each column, the best scores are highlighted with bold and underline (\under{best}), while the second-best scores are indicated in bold only (\textbf{second best}). Here \textbf{CrossGen}${}^*$ denotes the variation of our method in which the quad meshes are extracted using the same input mesh surfaces as for the other baseline methods, rather than using the predicted SDF-based surface (\textbf{CrossGen}).
    }
    \resizebox{0.98\linewidth}{!}{
    \begin{tabular}{l|cccccc}
    \hline\noalign{\smallskip}
      & Area~$\downarrow$ & Angle~$\downarrow$ & \# of Sings~$\downarrow$ & CD~$\downarrow$ & JR~$\uparrow$\\ 
    \noalign{\smallskip}
    \hline\noalign{\smallskip}
    MIQ~\footnotesize{~\cite{MIQ2009}} & 4.63 & 1.79 & \under{66.71} & 8.63 & 0.66\\
    Power Fields~\footnotesize{~\cite{Power_Fields2013}} & 4.55 & 1.77 & 85.41 & 8.76 & 0.72\\
    PolyVectors~\footnotesize{~\cite{PolyVectors_Diamanti2014}} & 4.53 & 1.72 & 87.21 & 8.75 & 0.71 \\
    IM~~\footnotesize{\cite{Instant_Meshes2015}} & 5.07 & 3.11 & 297.15 & 9.85 & 0.74\\
    QuadriFlow~~\footnotesize{\cite{QuadriFlow2018}} & 5.04 & 2.28 & 98.31 & 28.38 & 0.79\\
    NeurCross~~\footnotesize{\cite{NeurCross2024}} & \under{4.01} & \under{1.48} & \textbf{70.03} & \under{8.14} & \under{0.86}\\
    \hline
    \textbf{CrossGen (Ours)} & 4.32 & 1.60  & 80.65 & 8.45 & 0.83\\
    \textbf{\textbf{CrossGen}${}^*$ (Ours)} & \textbf{4.28} & \textbf{1.55}  & 78.17 & \textbf{8.23} & \textbf{0.84}\\
    \hline
    \end{tabular}
    \label{tab:comp_quad}
    }
\end{table}

\begin{figure*}[t]
  \centering
   \begin{overpic}[width=0.98\linewidth]{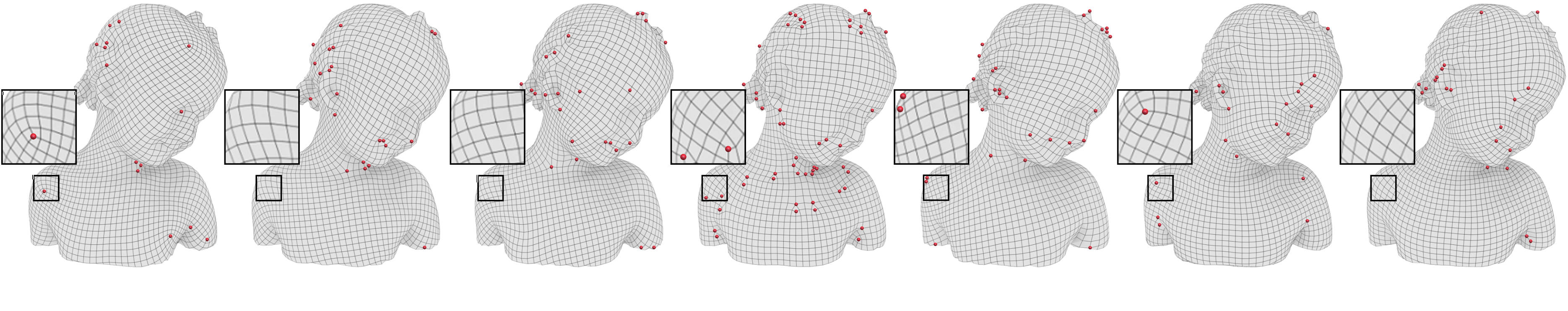}
        \put(2.8,-0.5){\small \begin{tabular}{c}
             33~\# of Sings  \\
             MIQ \\ 
        \end{tabular}}
        \put(17.3,-0.5){\small \begin{tabular}{c}
             45~\# of Sings  \\
             Power Fields \\ 
        \end{tabular}}
        \put(31.5,-0.5){\small \begin{tabular}{c}
             48~\# of Sings  \\
             PolyVectors \\ 
        \end{tabular}}
        \put(45,-0.5){\small \begin{tabular}{c}
             155~\# of Sings  \\
             IM \\ 
        \end{tabular}}
        \put(60,-0.5){\small \begin{tabular}{c}
             63~\# of Sings  \\
             QuadriFlow \\ 
        \end{tabular}}
        \put(74,-0.5){\small \begin{tabular}{c}
             38~\# of Sings  \\
             NeurCross \\ 
        \end{tabular}}
        \put(87,-0.5){\small \begin{tabular}{c}
             40~\# of Sings  \\
             $\text{CrossGen}^*$ (Ours) \\ 
        \end{tabular}}
   \end{overpic}
   \caption{
   Quad meshes generated by six baseline methods and our CrossGen. The red points mark the locations of singularities, where ``\# of Sings'' denotes the number of singular points in each quad mesh.
   Benefiting from large-scale data-driven learning, our method produces smoother and more regular quad meshes compared to all other approaches.
   }
   \label{fig:compare_quad}
\end{figure*}
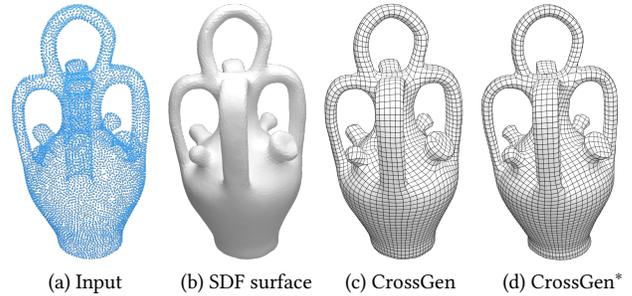
\begin{figure}[t]
  \centering
   \begin{overpic}[width=0.98\linewidth]{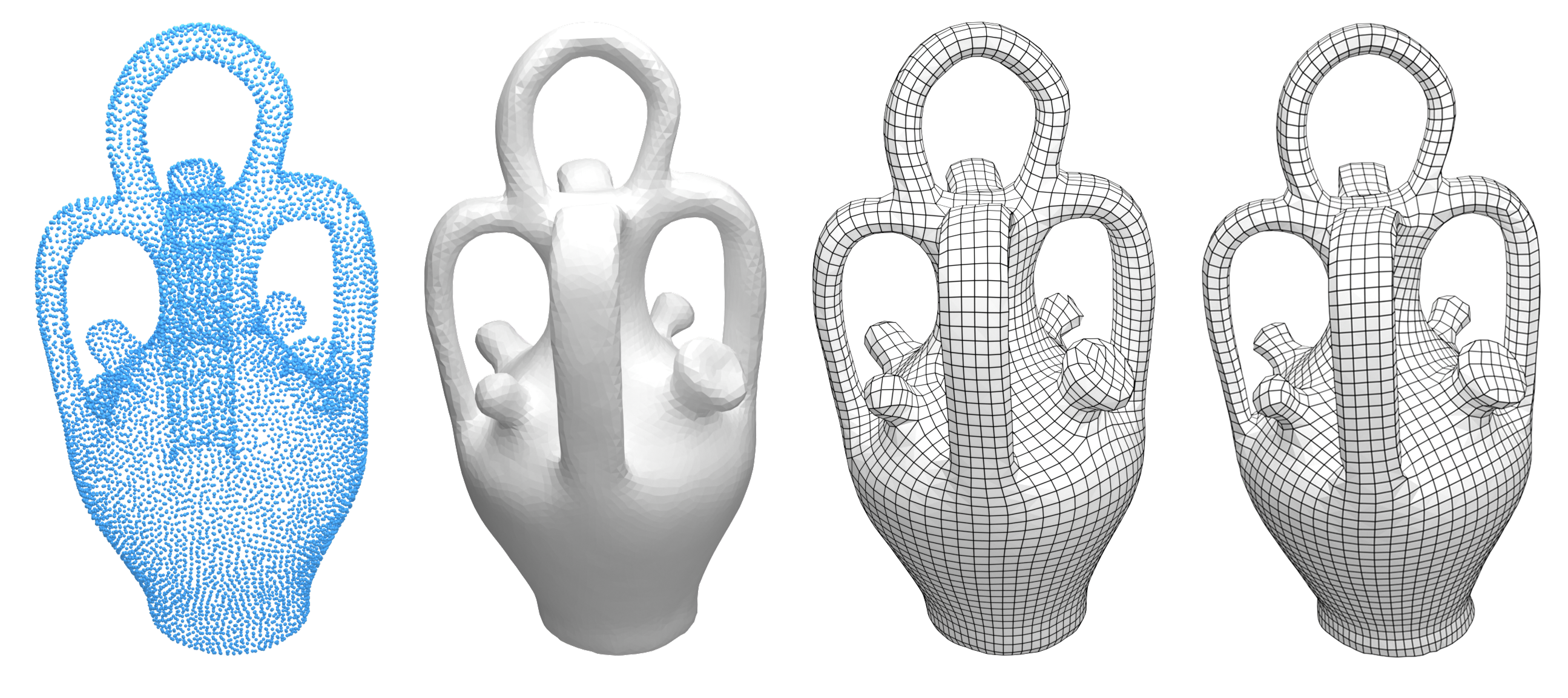}
        \put(8,-2){\small (a) Input}
        \put(29,-2){\small (b) SDF surface}     
        \put(53,-2){\small \begin{tabular}{c}
            (c) CrossGen
        \end{tabular}}
        \put(78,-2){\small \begin{tabular}{c}
             (d) $\text{CrossGen}^*$
        \end{tabular}}
   \end{overpic}
   \caption{
   SDF and quad mesh results from point cloud input. Since quad extraction requires triangular meshes, given a point-cloud surface (a), we use the SDF branch to reconstruct a triangle mesh (b), and extract quad mesh (c). (d) shows the extracted mesh from GT triangle mesh. 
   }
   \label{fig:input_points}
\end{figure}

\paragraph{Quad Mesh Evaluation}
To evaluate the effectiveness of our method in downstream applications, we apply it to quadrilateral meshing tasks and compare the resulting mesh quality with optimization-based approaches 
in Tab. \ref{tab:comp_quad}.
Except for IM~\cite{Instant_Meshes2015} and QuadriFlow~\cite{QuadriFlow2018}, all methods employ the same quad mesh extraction technique, libQEx~\cite{libQEX13}.
Note that our method assumes a point-cloud surface as input, whereas the existing quad mesh extraction methods requires a triangle mesh as the base surface for parameterization and quad extraction.
Since all the baseline methods operate on ground-truth triangle meshes, to evaluate our method comprehensively, we consider two settings in Tab. \ref{tab:comp_quad}:
(1) We test our model using the same ground-truth mesh as the base surface for querying the cross field and quad meshing, denoted as \textbf{CrossGen$^*$}.
(2) Starting from point cloud input, we use the SDF branch to reconstruct a surface via Marching Cubes~\cite{MC}, which is then used as the base surface for quad meshing, denoted as \textbf{CrossGen}.

Tab. \ref{tab:comp_quad} shows the quantitative results comparing with baseline methods. Our method ranks second only to NeurCross across multiple evaluation metrics, outperforming all other baselines.
We note that since NeurCross is used as the expert model to generate the GT data for training our \emph{CrossGen} network, its performance is expected to be an upper bound for \emph{CrossGen}’s performance.
Furthermore, although both our method and NeurCross~\cite{NeurCross2024} produce a greater number of singularities than MIQ~\cite{MIQ2009}, they achieve higher Jacobian Ratio~(JR) values, indicating that MIQ~\cite{MIQ2009} tends to generate more distorted quadrilaterals.
Fig.~\ref{fig:compare_quad} presents a visual comparison, where our method produces high-quality and structurally consistent quadrilateral meshes, with results that are overall better or comparable to those of all the baseline methods.
In Fig.~\ref{fig:compare_quad}, we also highlight the locations of singularities. The distribution of singular points in the quad mesh generated by our \emph{CrossGen} is similar to that produced by NeurCross~\cite{NeurCross2024}.

In addition, the setting of \textbf{CrossGen}, which reconstructs the surface from point cloud inputs via the SDF branch, achieves results comparable to \textbf{CrossGen$^*$}, which uses the GT mesh as input. 
A visual comparison of the two settings is shown in Fig.~\ref{fig:input_points}. These results highlight the effectiveness of our SDF branch in producing high-quality surface geometry suitable for downstream quad meshing tasks.

\begin{figure}
\centering
    \begin{overpic}[width=0.9\linewidth]{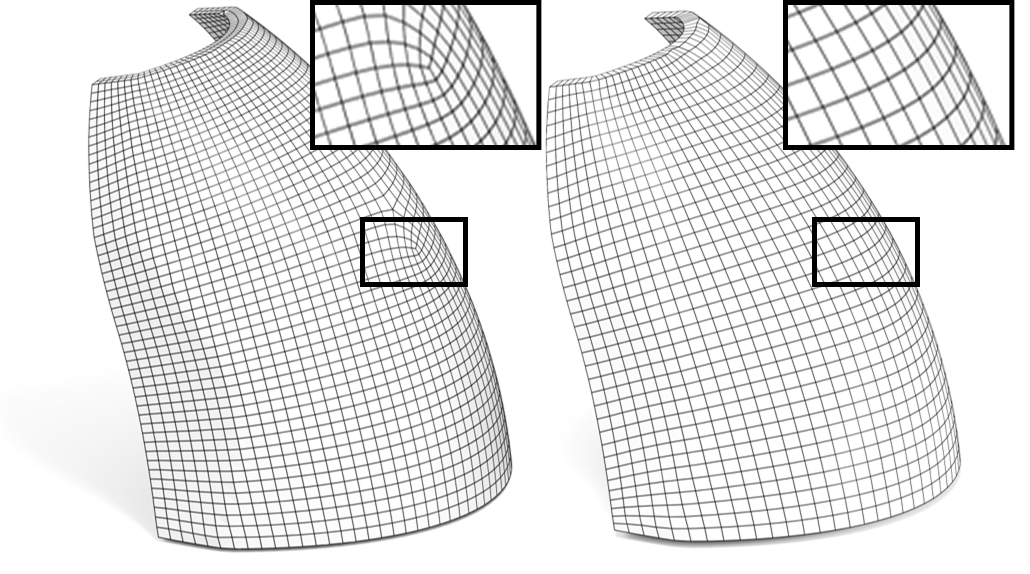}
        \put(22,-2){\small Point2Quad}
        \put(65,-2){\small CrossGen (Ours)}
    \end{overpic}
    \caption{
    Comparison with Point2Quad~\cite{li2025point2quad} on CAD shapes.
    Our method produces more regular, smoother, and principal curvature-aligned quad meshes, whereas Point2Quad often results in meshes with more singularities. The result for Point2Quad is obtained from its official code repository.
    }
    \label{fig:compare_point2quad}
\end{figure}

\paragraph{Comparison with Point2Quad} 
Point2Quad~\cite{li2025point2quad} is a learning-based method that generates quad meshes directly from point clouds, primarily targeting CAD-like geometries. In their paper, only results on CAD shapes are presented, and no pretrained models are publicly available. Therefore, for comparison, we refer to the qualitative results provided in their official repository. As shown in Fig.~\ref{fig:compare_point2quad}, while Point2Quad can generate plausible quads on simple CAD shapes, it struggles to preserve global field smoothness, resulting in increased singularities in the generated quad meshes. 
In contrast, our method produces field-aligned quad meshes with improved regularity. 

\begin{figure*}[t]
  \centering
   \begin{overpic}[width=0.98\linewidth]{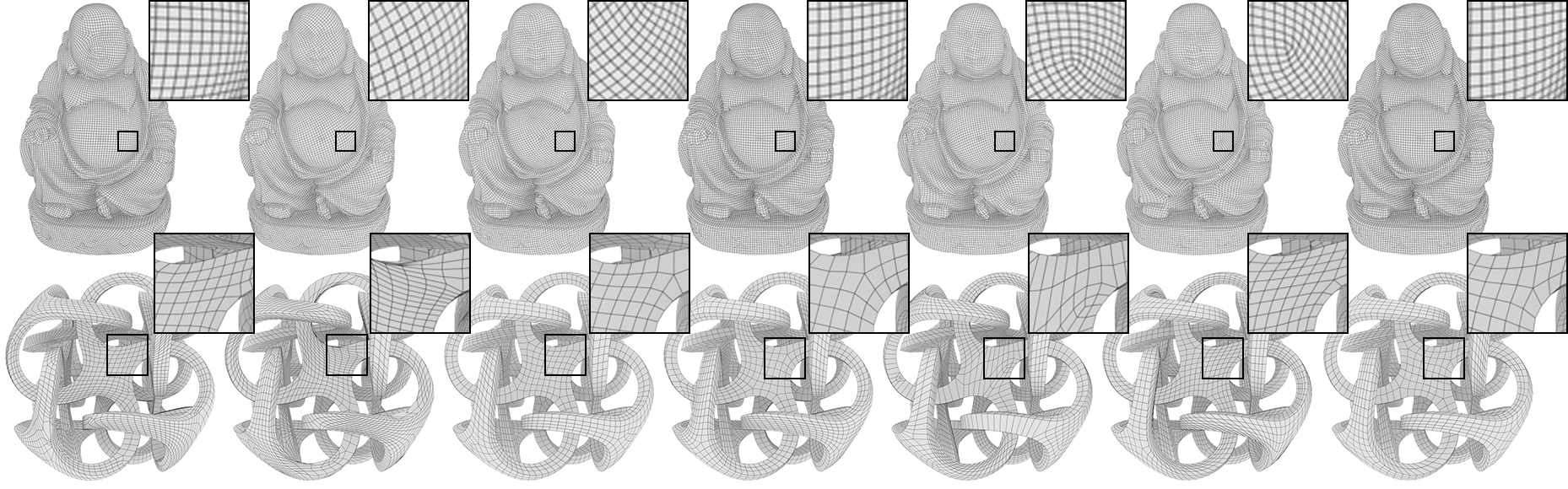}
        \put(5,-1.5){\small MIQ}
        \put(17,-1.5){\small Power Fields}
        \put(31,-1.5){\small PolyVectors}
        \put(47,-1.5){\small IM}
        \put(58,-1.5){\small QuadriFlow}
        \put(73,-1.5){\small NeurCross}
        \put(86,-1.5){\small $\text{CrossGen}^*$ (Ours)}
        \put(-2,21){\small \rotatebox{90}{libQEx}}
        \put(-2,3){\small \rotatebox{90}{QuadWild}}
   \end{overpic}
   \caption{
   Visualization of quad meshes generated for a free-form shape and a CAD-type shape by six baseline methods and our CrossGen. The free-form quad mesh is extracted using libQEx~\cite{libQEX13}, while the CAD-type quad mesh is extracted using QuadWild~\cite{quadwild2021}.
   }
   \label{fig:extractor}
\end{figure*}
\begin{figure}
    \centering
\begin{overpic}[width=0.98\linewidth]{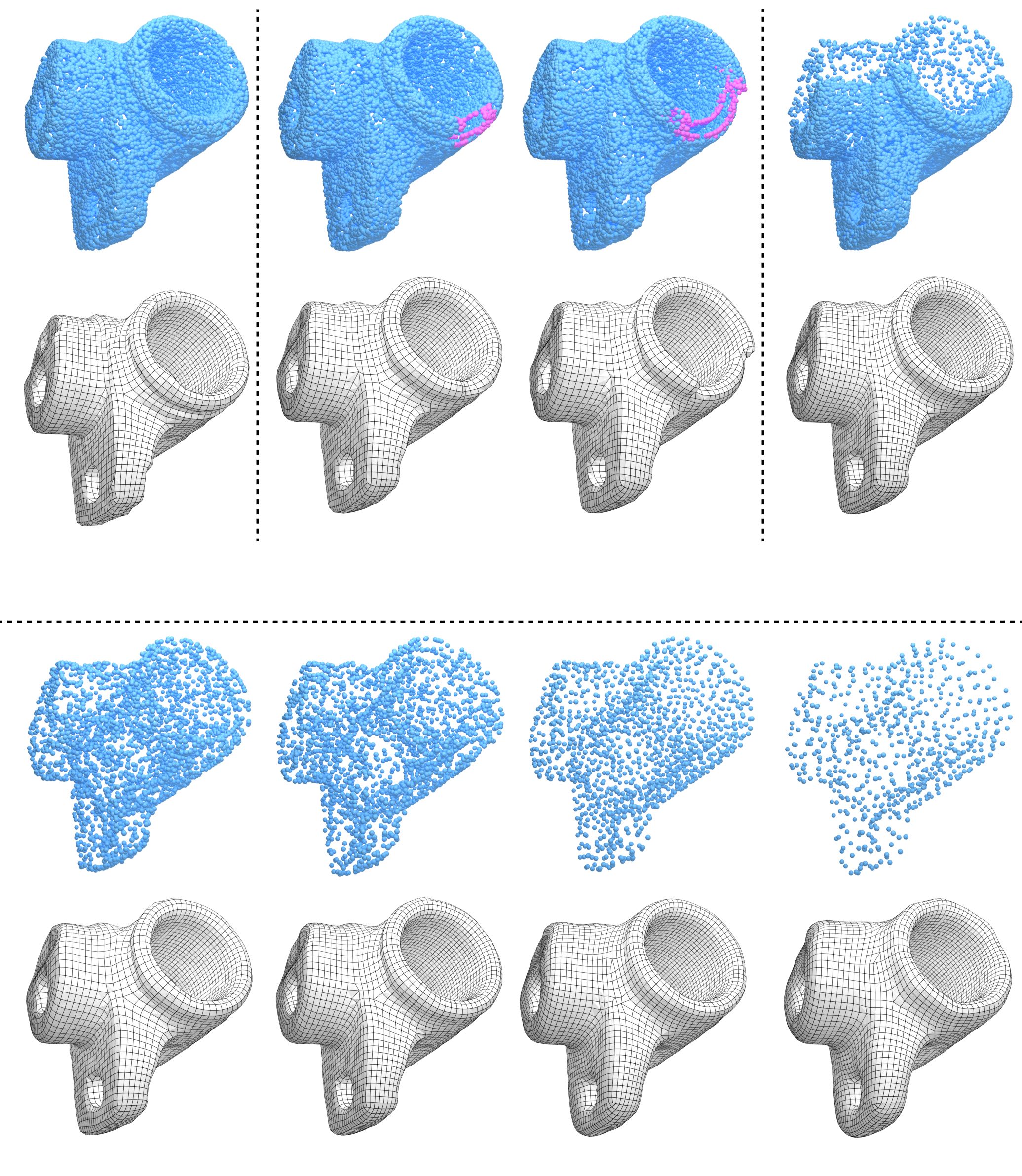}
    \put(0,50){\small (a) Original shape}
    \put(28,50){\small (b) Point missing shapes}
    \put(66,50){\small (c) Uneven input}
    \put(-2,10){\small \rotatebox{90}{(d) Sparse point shape}}
    \put(3,-1){\small 10K points}
    \put(25,-1){\small 5K points}
    \put(46,-1){\small 3K points}
    \put(69,-1){\small 1K points}
    \end{overpic}
    \caption{
    Quad meshes generated by our method under varying input conditions, where the normal of our input point cloud is obtained using~\citet{normal_Xu506} method.
    (a) the original point cloud with 150K points; (b) a point cloud with missing regions (highlighted in light violet); (c)  an uneven point cloud input; and (d) Point cloud inputs with varying levels of sparsity.
    }
    \label{fig:missing_points_and_sparse_sample}
\end{figure}

\paragraph{Extraction Methods}

We use existing methods to generate quad meshes guided by cross fields produced by \emph{CrossGen}. There three widely adopted open-source extractors: the IM method~\cite{Instant_Meshes2015}, libQEx~\cite{libQEX13}, and QuadWild~\cite{quadwild2021}. The IM extractor, specifically designed for IM~\cite{Instant_Meshes2015}, leverages local parameterization to accelerate the extraction process. While it offers high efficiency, this speed often comes at the expense of mesh quality, typically resulting in increased singularities and non-quad face (e.g, triangle faces as shown in Fig.~\ref{fig:compare_quad}). 
In comparison, libQEx~\cite{libQEX13} and QuadWild~\cite{quadwild2021} tend to generate higher-quality quad meshes, though they are slower than IM. As noted in NeurCross~\cite{NeurCross2024}, libQEx~\cite{libQEX13} is particularly suited for quad meshing of free-form shapes, while QuadWild~\cite{quadwild2021} is specifically designed to preserve sharp features and thus well-suited for CAD models and surfaces with open boundaries. Hence, we will test the performance of \emph{CrossGen} in combination with both libQEx and QuadWild in the quad meshing task.

Fig.~\ref{fig:extractor} shows quad meshes generated using libQEx~\cite{libQEX13} and QuadWild~\cite{quadwild2021}, respectively, each driven by cross fields predicted by different baseline methods, as well as \emph{CrossGen}. With each mesh extraction method fixed, the mesh quality is solely determined by the quality of the cross field provided. Fig.~\ref{fig:extractor} shows that \emph{CrossGen} consistently produces quad meshes of better quality than other baseline methods, indicating the superior quality of the cross field predicted by \emph{CrossGen}. 

Although \emph{CrossGen} delivers extremely fast computation of cross fields, we have not accelerated the step of quad mesh extraction.  
Specifically, on our test dataset, libQEx~\cite{libQEX13} and QuadWild~\cite{quadwild2021} require an average of 40.81 seconds and 15.17 seconds, respectively, to extract a quad mesh from a given cross fields.
Since the inference time for cross field prediction in our method is negligible, the total runtime from the input point cloud to the final quad mesh ranges from 15 to 41 seconds, depending on whether QuadWild~\cite{quadwild2021} or libQEx~\cite{libQEX13} is used to extract the quad mesh.

\subsection{Analysis}
\label{sec:futher_comparison}

\paragraph{Robustness to Missing, Uneven, and Sparse Inputs.}
A key strength of \emph{CrossGen} lies in its robustness to diverse point cloud inputs, enabled by the flexibility of using point clouds. Three common scenarios are incomplete, uneven, and sparsely sampled data. In real-world settings, 3D scans frequently suffer from missing regions, uneven sampling, or low-resolution sampling. Unlike traditional methods that rely on complete surface meshes, \emph{CrossGen} remains effective even when substantial portions of the input geometry are absent. By leveraging a locality-aware latent space, our model can infer global structure and cross fields from limited data.
As illustrated in Fig.~\ref{fig:missing_points_and_sparse_sample}, \emph{CrossGen} successfully reconstructs coherent surfaces and high-quality quadrilateral meshes under challenging conditions, including inputs with missing regions (b), uneven sampling (c), and varying degrees of sparsity (d). We note, however, that excessively sparse inputs may result in the loss of detail, leading to overly smooth reconstructions.

\begin{figure}
    \centering
\begin{overpic}[width=\linewidth]{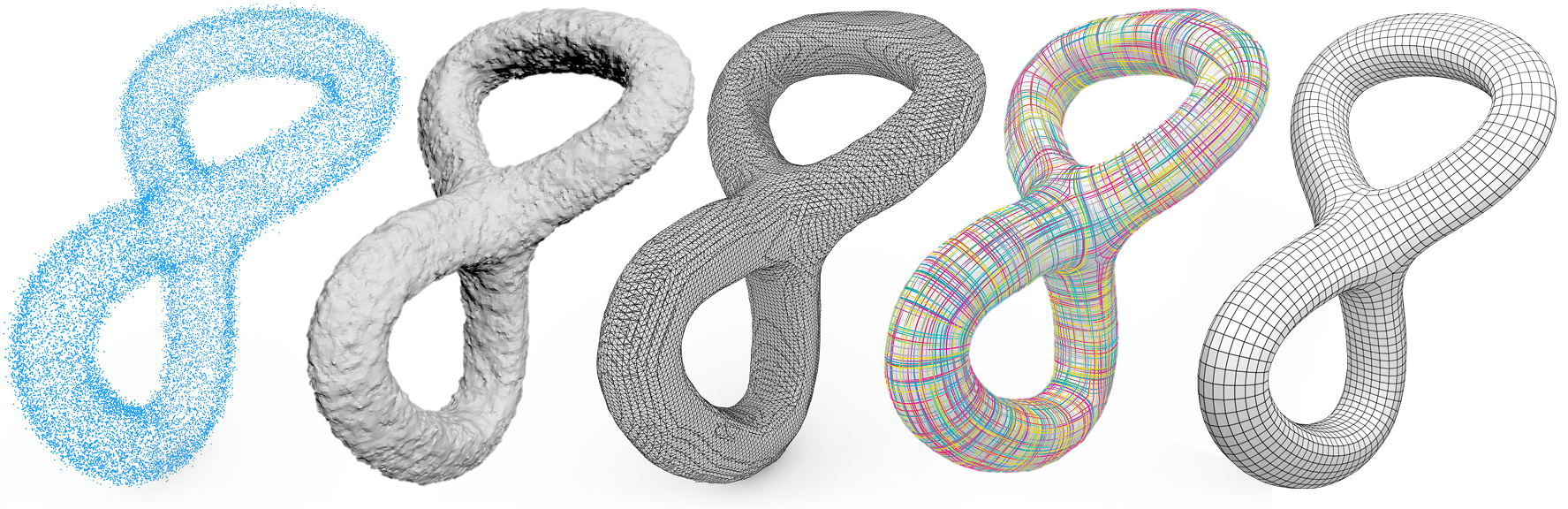}
    \put(7,-3){(a)}
    \put(26,-3){(b)}
    \put(45,-3){(c)}
    \put(65,-3){(d)}
    \put(83,-3){(e)}
\end{overpic}
    \caption{
    Cross field generation from noisy point cloud input. Given noisy input (a), Poisson reconstruction produces a noisy surface (b), whereas our SDF branch reconstructs a smooth geometry (c). Surface points sampled from our geometry enable accurate cross field prediction (d) and subsequent quad meshing (e), demonstrating robust cross field generation from noisy data.
    }
    \label{fig:pointcloud2cf}
\end{figure}
\begin{figure}
    \centering
\begin{overpic}[width=0.9\linewidth]{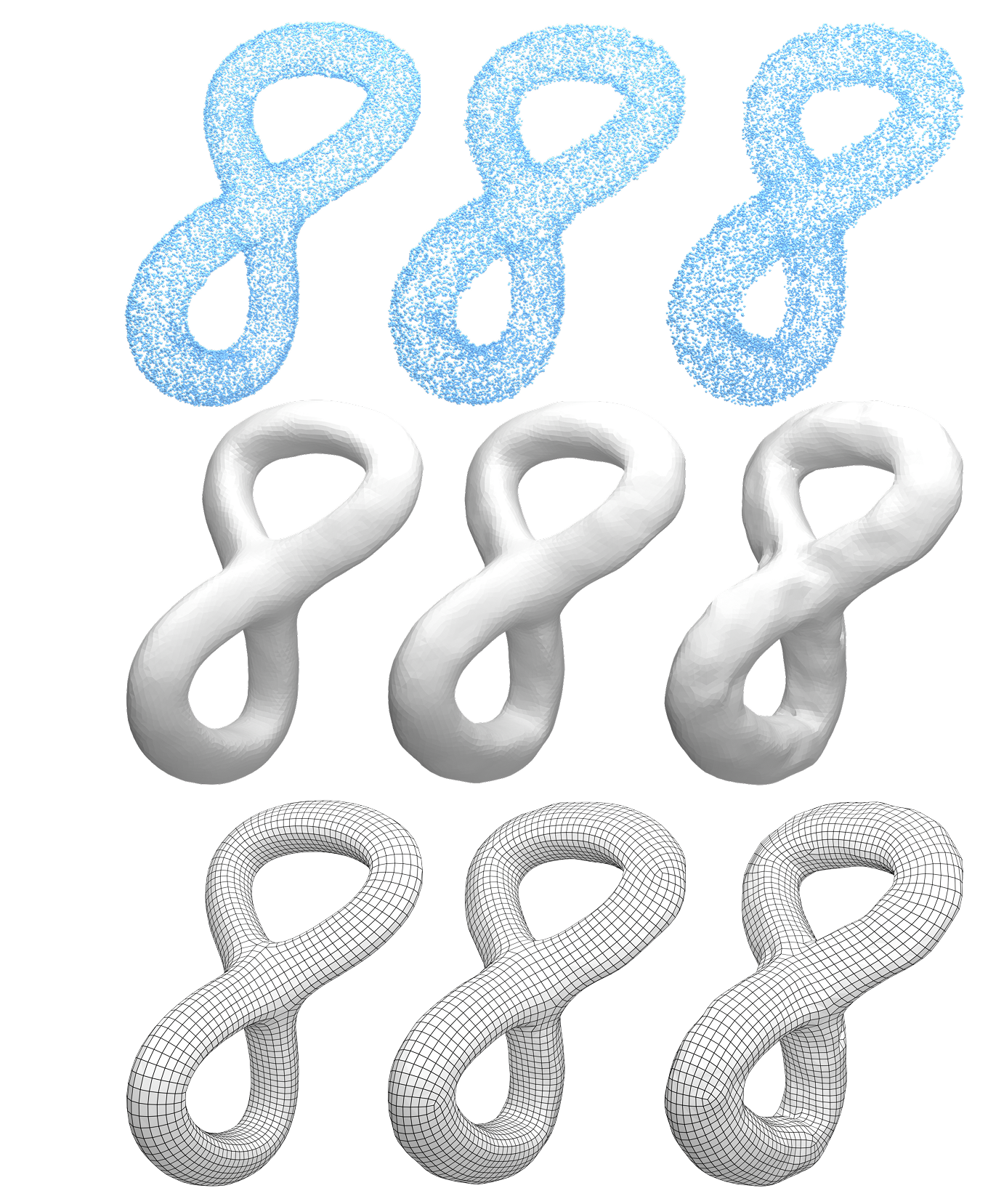}
    \put(5,-2){\small 1\% Gaussian noise}
    \put(30,-2){\small 2\% Gaussian noise}
    \put(55,-2){\small 3\% Gaussian noise}

     \put(1,83){\small Noise input}
     \put(1,52){\small SDF surface}
     \put(1,19){\small Quad Mesh}
    
    \end{overpic}
    \caption{
    To evaluate noise robustness, we test our method on point clouds with varying noise levels, where the normals of the input point cloud are estimated using~\citet{normal_Xu506} method.
    From left to right: As noise levels increase, the SDF surface and quad mesh quality degrade moderately, yet the overall quad mesh structure remains consistent, with only a slight rise in singularities.
    }
    \label{fig:noise}
\end{figure}
\begin{figure}
    \centering
    \begin{overpic}[width=\linewidth]{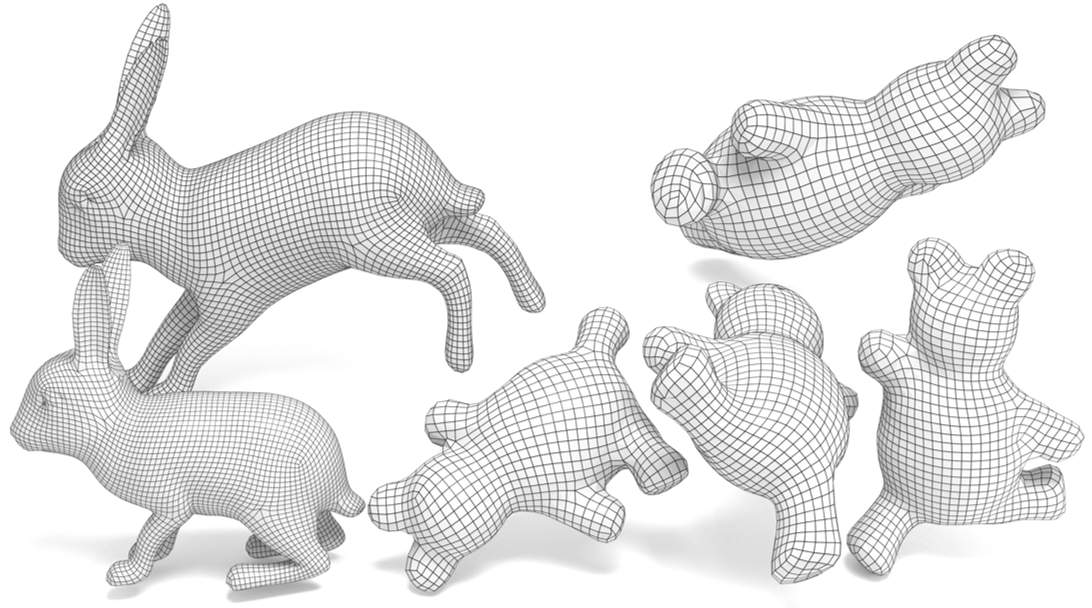}
  \end{overpic}
    \caption{
    Visualization of quad meshes generated by our CrossGen under random poses, including both non-rigid deformation and rigid rotation.
    }
    \label{fig:rotation}
\end{figure}

\paragraph{Resistance to Noise}
For point cloud inputs (Fig.~\ref{fig:pointcloud2cf}~(a)), a triangle mesh surface is required to produce a quad mesh. One option is to use Poisson reconstruction~\cite{SPSR2013} to generate the surface (Fig.~\ref{fig:pointcloud2cf}~(b)), but this approach can struggle to produce accurate results especially when the point cloud contains significant noise. In contrast, our method learns both SDF and cross field representations, enabling us to directly infer a high-quality SDF output from point cloud data via the SDF branch, even for noisy point clouds. We then extract the geometry surface using the Marching Cubes algorithm~\cite{MC} (Fig.~\ref{fig:pointcloud2cf}~(c)), sample points on the surface, and query the decoder’s feature volume to generate the corresponding cross field (Fig.~\ref{fig:pointcloud2cf}~(d)) and extract the final quad mesh (Fig.~\ref{fig:pointcloud2cf}~(e)). 
In Fig.~\ref{fig:noise}, we further evaluate the robustness of our model under varying levels of Gaussian noise. Despite increasing noise, our method consistently produces smooth surfaces and coherent quad meshes, demonstrating strong resilience to input perturbations.

\paragraph{Robustness to Rotation} 

Another important feature of \emph{CrossGen} is its robustness to random rotations. 
Data-driven methods, such as~\citet{DL2quadMesh2021}, often require shapes to be in a canonical pose. In contrast, by leveraging a learned local latent space representation and data augmentations, \emph{CrossGen} can handle arbitrary orientations, making it robust to input pose variations such as rotations and non-rigid deformations, as illustrated in Fig.~\ref{fig:rotation}.

\begin{figure}
    \centering
    \begin{overpic}[width=\linewidth]{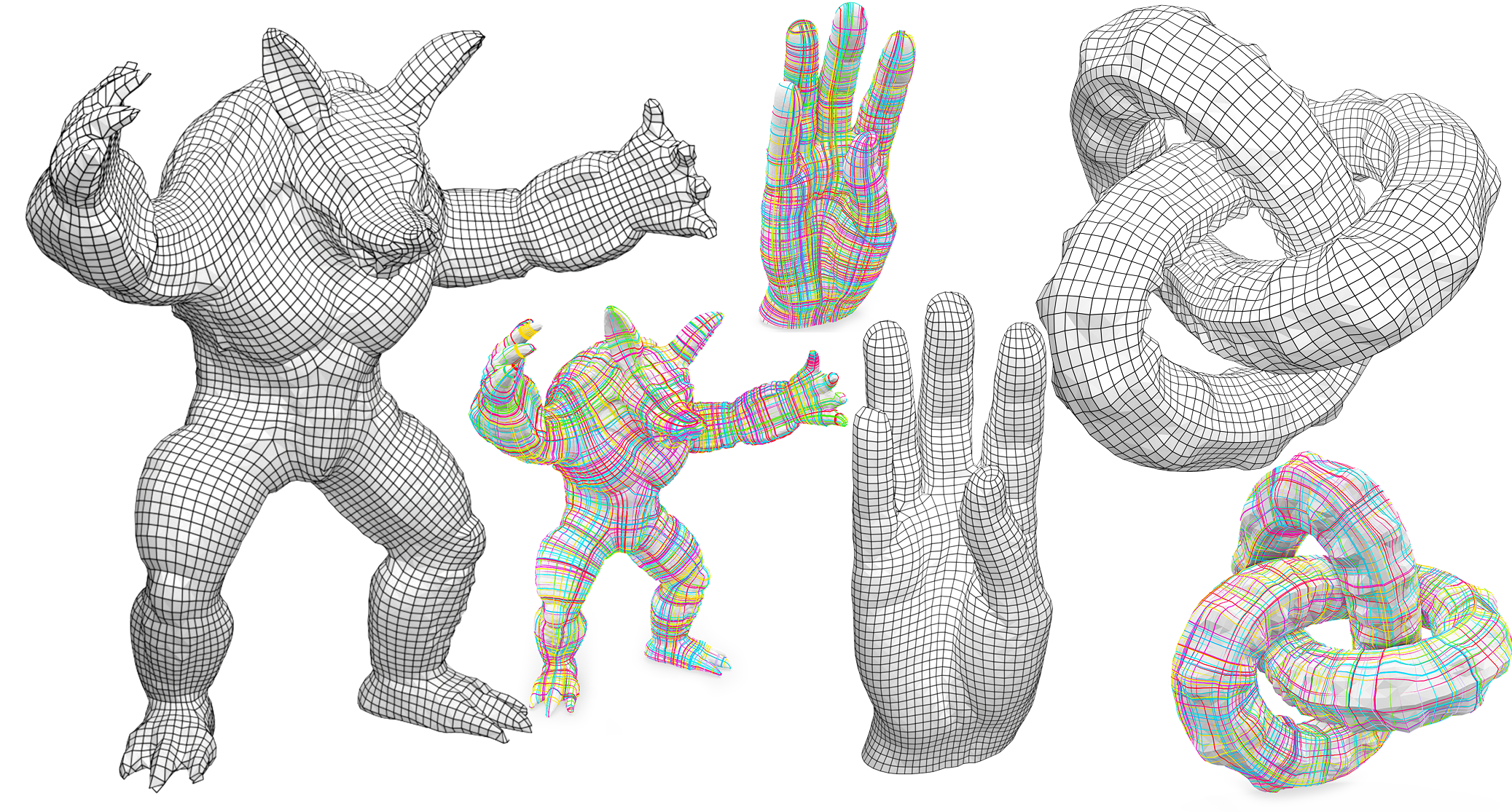}
  \end{overpic}
    \caption{
    Out-of-domain generalization of our CrossGen on unseen shape categories.
    }
    \label{fig:derversity}
\end{figure}

\paragraph{Generalization to Out-of-domain Shapes} 
One more key strength of \emph{CrossGen} is its ability to handle out-of-domain geometries and new shape categories.
In our experiments, we tested our method on shapes from previously unseen categories in the training stage.
Despite not being trained on these specific shapes, \emph{CrossGen} successfully predicts high-quality cross fields, demonstrating its ability to generalize across a wide range of geometries and surface types, as shown in Fig. \ref{fig:derversity}. This ability is crucial for real-world applications where new, previously unseen shapes may need to be processed without requiring retraining or fine-tuning on each new dataset. 

\begin{figure}
    \centering
    \begin{overpic}[width=\linewidth]{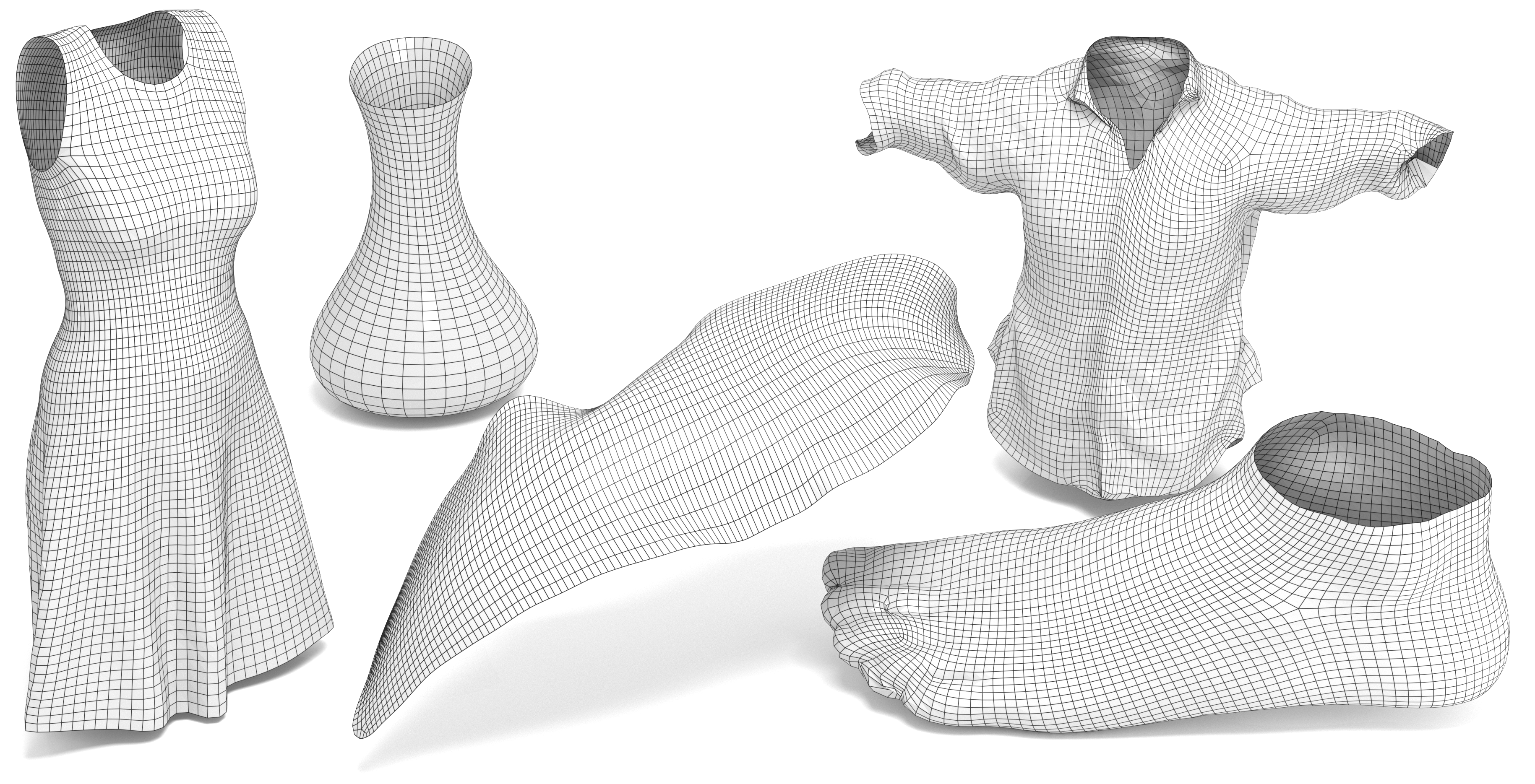}
    \end{overpic}
    \vspace{-2mm}
    \caption{
    Cross field generation on surfaces with open boundaries by our CrossGen.
    Our method robustly generalizes to open-boundary surfaces, enabling high-quality quad mesh extraction from the predicted cross fields.
    }
    \label{fig:openbound2cf}
\end{figure}
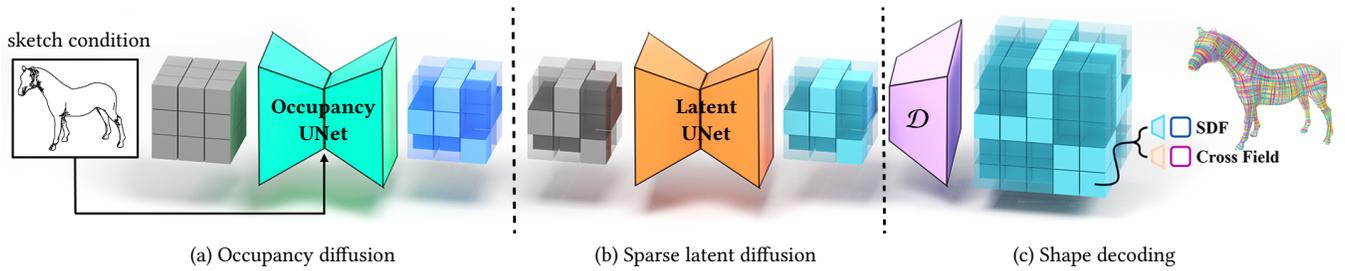
\begin{figure*}
    \centering
    \begin{overpic}[width=\linewidth]{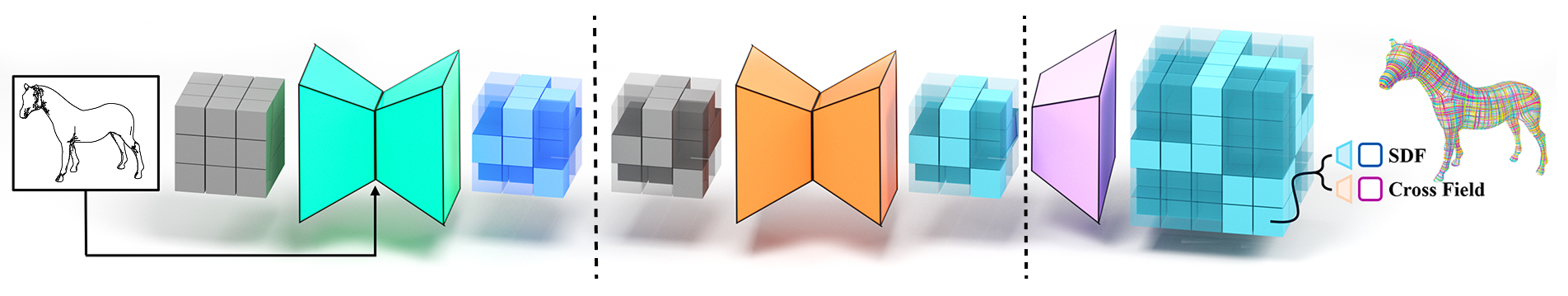}
        \put(14,-1){\small (a) Occupancy diffusion}
        \put(44,-1){\small (b) Sparse latent diffusion}
        \put(75,-1){\small (c) Shape decoding}

        \put(19.7,15.5){\small \textbf{\begin{tabular}{c}
             Occupancy  \\
             UNet \\ 
        \end{tabular}}}
        \put(49.3,15.3){\small \textbf{\begin{tabular}{c}
             Latent  \\
             UNet \\ 
        \end{tabular}}}

        \put(0,15){ \small sketch condition}
        \put(67, 9){\large $\mathcal{D}$}
    \end{overpic}
    \caption{
    The sketch-conditioned diffusion pipeline in our CrossGen, following LAS-Diffusion~\cite{zheng2023LASDiffusion}.  (a) Occupancy diffusion conditioned on the input sketch: predicts coarse voxel structure to capture global shape layout from dense voxel noise. (b) Sparse latent diffusion: refines geometry and cross field information from sparse latent noise in the latent space . (c) Shape decoding: converts denoised latent codes into a 3D shape and corresponding cross field via our pretrained decoder.
    }
    \label{fig:method_overview_diff_gen}
\end{figure*}

\paragraph{Open-boundary Surfaces}
Given a set of 3D points sampled from open-boundary surfaces, we first encode them into a high-resolution feature grid, then query the corresponding cross fields via the cross field branch, and finally extract the quad mesh using QuadWild~\cite{quadwild2021}.
Fig.~\ref{fig:openbound2cf} presents quad mesh results generated by our method on the shapes with open boundaries.
Although our training dataset does not explicitly include surfaces with open boundaries, we find that our model generalizes well to such surfaces. 
We attribute this capability to the use of a sparse and local encoder in our model’s encoding process. By focusing on local surface features, the encoder inherently emphasizes learning local geometry, enabling greater flexibility and generalization.

\subsection{Quad Meshing of Novel Generated Shapes} 
\label{sec:Generated_Shapes}

To take advantage of the latent space learned by \emph{CrossGen}, we have also explored the extension of \emph{CrossGen} to generative modeling. Specifically, we design a two-stage diffusion pipeline that operates directly within this latent space, inspired by LAS-Diffusion~\cite{zheng2023LASDiffusion} and TRELLIS~\cite{xiang2024TRELLIS}. As illustrated in Fig.~\ref{fig:method_overview_diff_gen}, our quad mesh generation pipeline begins with an \emph{occupancy diffusion model} on sketch conditions (Fig.~\ref{fig:method_overview_diff_gen} (a)), which synthesizes a coarse voxel occupancy distribution in the latent space, providing a rough approximation of the global structure and topology of the target shape. In the second stage, conditioned on the coarse occupancy, a \emph{fine-grained latent diffusion model} refines the representation by generating high-resolution latent embeddings that capture detailed geometry and cross field information necessary for downstream meshing (Fig.~\ref{fig:method_overview_diff_gen} (b)).
Finally, the resulting latent codes are decoded into a cross field by the decoder~$\mathcal{D}$ (Fig.~\ref{fig:method_overview_diff_gen} (c)).

\begin{figure}
    \centering
    \begin{overpic}[width=\linewidth]{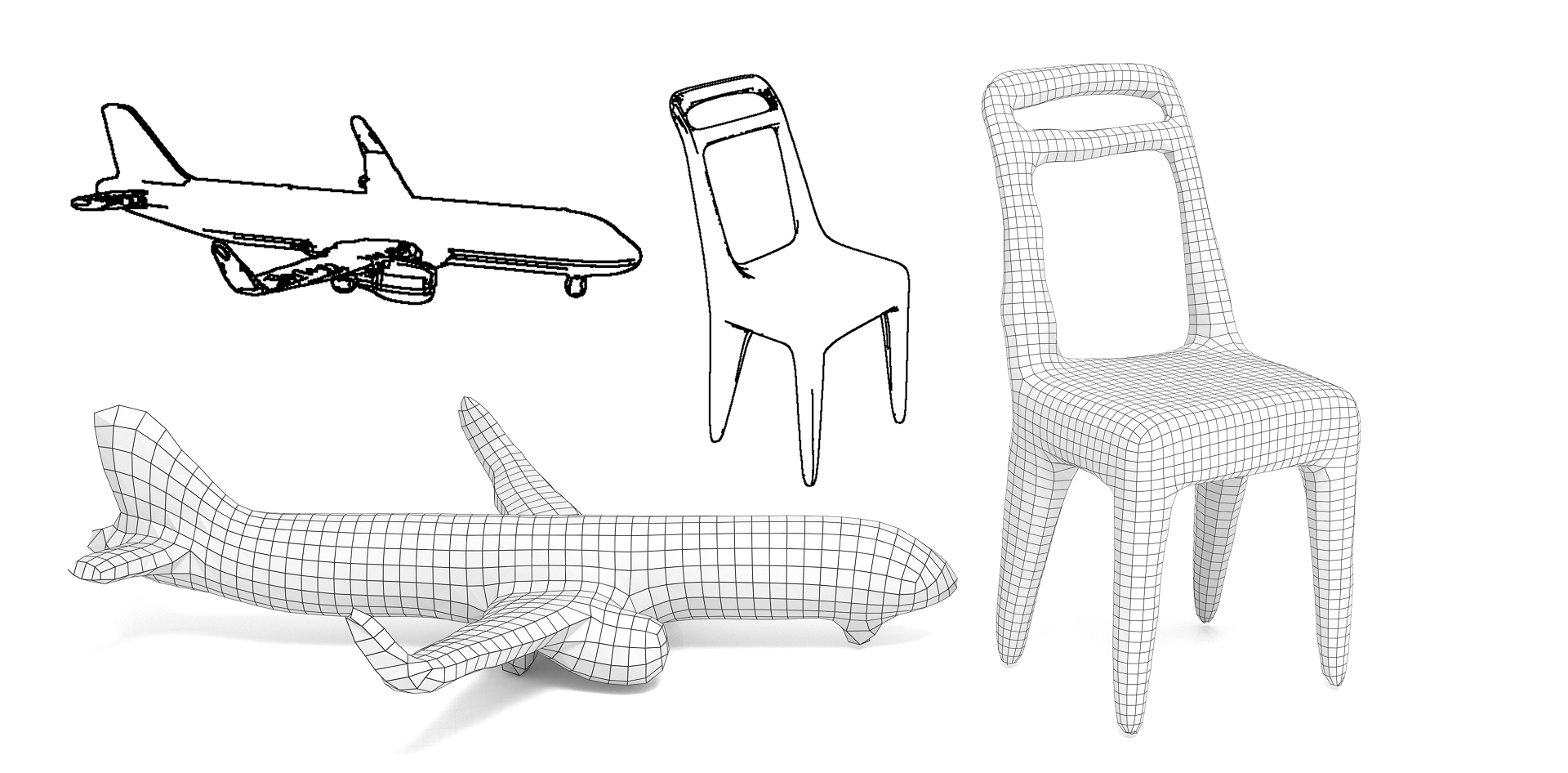}
    \end{overpic}
    \vspace{-6mm}
    \caption{
    The quad meshes are generated by our CrossGen via sketch-conditioned diffusion.
    }
    \label{fig:vis_diff_gen}
\end{figure}

To demonstrate the effectiveness of generative capacity, we train the proposed two-stage diffusion pipeline on the full chair and airplane categories. 
Our method successfully generates cross fields that produce diverse and coherent quad meshes, consistently aligned with the input sketch conditions, as shown in Fig.~\ref{fig:vis_diff_gen}.
The results show that our latent space captures both global shape priors and cross field details necessary for high-quality quad meshing.

\section{Ablation Study}
\begin{figure}
    \centering
    \begin{overpic}[width=0.8\linewidth]{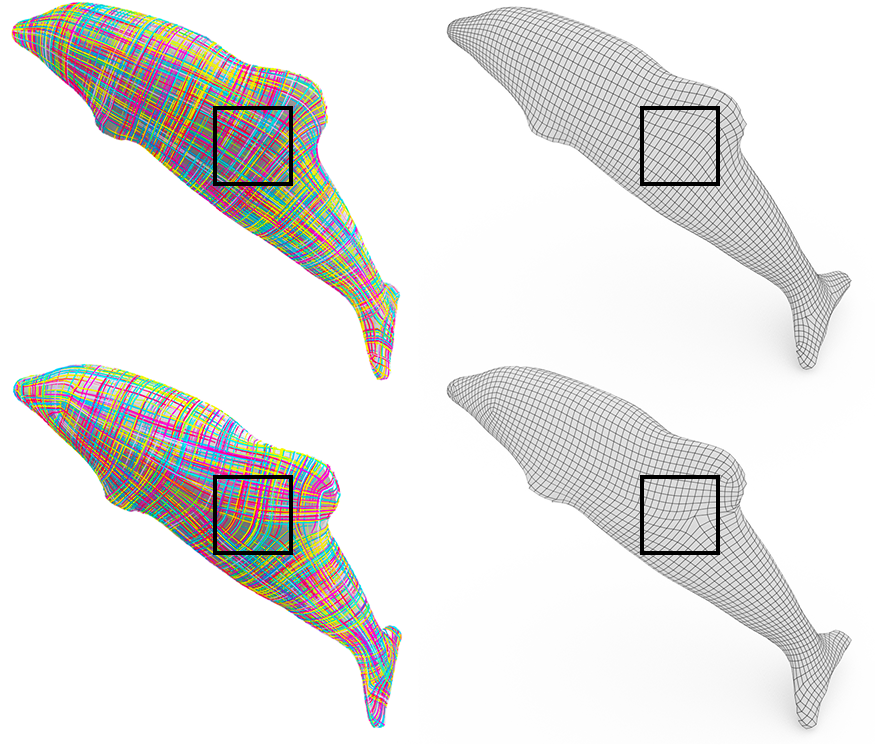}
        \put(-12, 55){\small (a) Predicted cross field}
        \put(-12,10){\small (b) Predicted rotation angle}
        
    \end{overpic}
    \caption{
    Ablation on two cross field representations: (a) direct prediction of cross field direction (Ours); (b) prediction of the rotation angle based on the local coordinate system. Our method yields more consistent predictions.
    }
    \label{fig:ablation_cf}
\end{figure}

To demonstrate the effectiveness of \emph{CrossGen}, we conduct an ablation study to compare two different cross field representations. 
Existing methods, such as those proposed in \citet{MIQ2009}, typically initialize a local coordinate system, predict the rotation angle of this coordinate system, and then generate a smooth cross field. In contrast, our approach simplifies this process by directly predicting one of the two cross field directions, which is perpendicular to the surface normal. The second direction is then calculated via the cross product. We compare these two representations while keeping the rest of the network architecture constant to ensure fairness using the metric AE (Sec.~\ref{sec:metrics}). Tab.~\ref{tab:ablation_study} summarizes the prediction results, showing that our method significantly outperforms the commonly used representation. Note that our method can more easily ensure the overall consistent cross field, as illustrated in Fig.~\ref{fig:ablation_cf}.

\begin{table}
    \centering
    \caption{
    Ablation study on cross field representations. We compare two prediction strategies: (a) ours, predicting one cross field direction, and (b) rotation angle prediction. Our direction prediction strategy achieves the best performance.
    }
    \begin{tabular}{cccc}
    \hline\noalign{\smallskip}
    & Angular Error (AE)~$\downarrow$ & Time (ms)~$\downarrow$ \\ 
    \noalign{\smallskip}
    \hline\noalign{\smallskip}
    Rotation & 0.34 &  73.26 \\
    Ours & \textbf{0.05} & \textbf{66.75} \\
    \hline
    \end{tabular}

    \label{tab:ablation_study}
\end{table}

\section{Limitations}
\begin{figure}
    \centering
    \begin{overpic}[width=\linewidth]{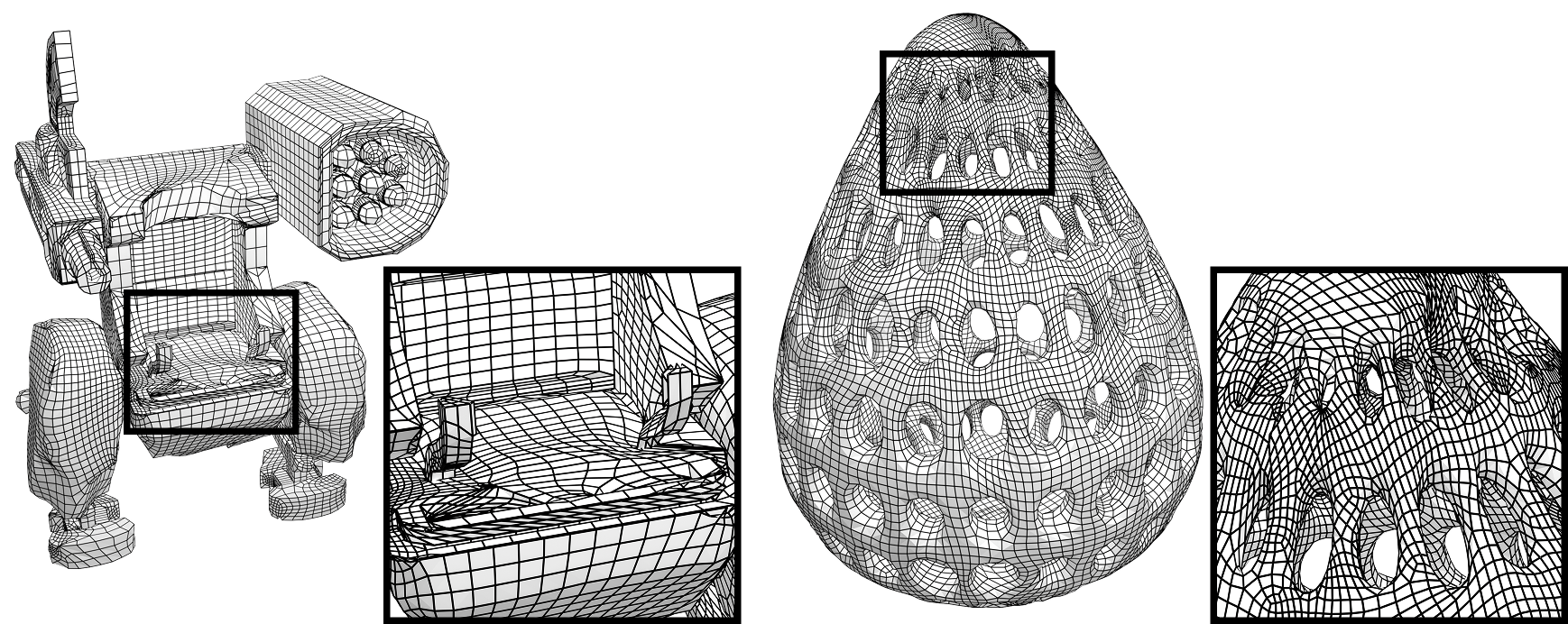}
    \end{overpic}
    \vspace{-6mm}
    \caption{Failure case. Our method struggles in the presence of fine-grained geometric details, resulting in quad mesh that lack smoothness and global consistency.}
    \label{fig:limitation}
\end{figure}

There are three limitations that should be addressed in future work. First, our current training set is relatively small compared to large foundational models used in the 3D AIGC field, which may limit the generalization capabilities of our model for rare geometries. 
Second, while \emph{CrossGen} works well for many types of shapes, it struggles with processing very complex objects as illustrated by the failure case in Fig.~\ref{fig:limitation}.
These fine-grained details lead to distorted quad meshes, posing challenges for accurate cross field prediction.
Finally, in this work, we take an initial step toward exploring the use of the learned latent space for novel quad mesh generation. Scaling up this approach to large-scale datasets, such as Objaverse \cite{deitke2023objaverse}, remains an open challenge, primarily due to the lack of high-quality ground-truth quad meshes, and is left for future work.

\section{Conclusion}
\emph{CrossGen} provides a novel framework for cross field generation directly from point clouds. By leveraging a joint latent space representation, our method enables efficient and robust decoding of both geometry and cross field information from a shared latent space, allowing for real-time cross field prediction without reliance on iterative optimization. The model exhibits strong generalization capabilities, effectively handling diverse geometric categories, including smooth organic shapes, CAD models, and surfaces with open boundaries, even when such cases were absent from the training data.

\begin{acks}
The authors would like to thank the anonymous reviewers for their valuable comments and suggestions. 
This work was partially supported by the Innovation and Technology Commission of the HKSAR Government under the InnoHK initiative,  the Hong Kong RGC (Ref. T45-205/21-N),  the National Key R\&D Program of China (2022YFB3303200), the National Natural Science Foundation of China (U23A20312, 62272277), the German Research Foundation within the Gottfried Wilhelm Leibniz program, and the National Science Foundation (OAC-1845962, OAC-1910469, and OAC-2311245).
\end{acks}

\bibliographystyle{ACM-Reference-Format}
\bibliography{main}


\begin{thebibliography}{51}


\ifx \showCODEN    \undefined \def \showCODEN     #1{\unskip}     \fi
\ifx \showDOI      \undefined \def \showDOI       #1{#1}\fi
\ifx \showISBNx    \undefined \def \showISBNx     #1{\unskip}     \fi
\ifx \showISBNxiii \undefined \def \showISBNxiii  #1{\unskip}     \fi
\ifx \showISSN     \undefined \def \showISSN      #1{\unskip}     \fi
\ifx \showLCCN     \undefined \def \showLCCN      #1{\unskip}     \fi
\ifx \shownote     \undefined \def \shownote      #1{#1}          \fi
\ifx \showarticletitle \undefined \def \showarticletitle #1{#1}   \fi
\ifx \showURL      \undefined \def \showURL       {\relax}        \fi
\providecommand\bibfield[2]{#2}
\providecommand\bibinfo[2]{#2}
\providecommand\natexlab[1]{#1}
\providecommand\showeprint[2][]{arXiv:#2}

\bibitem[Beaufort et~al\mbox{.}(2017)]%
        {beaufort2017computing}
\bibfield{author}{\bibinfo{person}{Pierre-Alexandre Beaufort}, \bibinfo{person}{Jonathan Lambrechts}, \bibinfo{person}{Fran{\c{c}}ois Henrotte}, \bibinfo{person}{Christophe Geuzaine}, {and} \bibinfo{person}{Jean-Fran{\c{c}}ois Remacle}.} \bibinfo{year}{2017}\natexlab{}.
\newblock \showarticletitle{Computing cross fields A PDE approach based on the Ginzburg-Landau theory}.
\newblock \bibinfo{journal}{\emph{Procedia engineering}}  \bibinfo{volume}{203} (\bibinfo{year}{2017}), \bibinfo{pages}{219--231}.
\newblock


\bibitem[Bommes et~al\mbox{.}(2013b)]%
        {IGM2013}
\bibfield{author}{\bibinfo{person}{David Bommes}, \bibinfo{person}{Marcel Campen}, \bibinfo{person}{Hans-Christian Ebke}, {et~al\mbox{.}}} \bibinfo{year}{2013}\natexlab{b}.
\newblock \showarticletitle{Integer-Grid Maps for Reliable Quad Meshing}.
\newblock \bibinfo{journal}{\emph{ACM Transactions on Graphics}} \bibinfo{volume}{32}, \bibinfo{number}{4} (\bibinfo{year}{2013}), \bibinfo{pages}{1--12}.
\newblock


\bibitem[Bommes et~al\mbox{.}(2013a)]%
        {BCE13_quad}
\bibfield{author}{\bibinfo{person}{D. Bommes}, \bibinfo{person}{M. Campen}, \bibinfo{person}{H.-C. Ebke}, \bibinfo{person}{P. Alliez}, {and} \bibinfo{person}{L. Kobbelt}.} \bibinfo{year}{2013}\natexlab{a}.
\newblock \showarticletitle{Integer-grid maps for reliable quad meshing}.
\newblock \bibinfo{journal}{\emph{ACM Transactions on Graphics (TOG)}} \bibinfo{volume}{32}, \bibinfo{number}{4} (\bibinfo{year}{2013}), \bibinfo{pages}{98}.
\newblock


\bibitem[Bommes et~al\mbox{.}(2009)]%
        {MIQ2009}
\bibfield{author}{\bibinfo{person}{David Bommes}, \bibinfo{person}{Henrik Zimmer}, {and} \bibinfo{person}{Leif Kobbelt}.} \bibinfo{year}{2009}\natexlab{}.
\newblock \showarticletitle{Mixed-Integer Quadrangulation}.
\newblock \bibinfo{journal}{\emph{ACM Transactions on Graphics}} \bibinfo{volume}{28}, \bibinfo{number}{3} (\bibinfo{year}{2009}), \bibinfo{pages}{1--10}.
\newblock


\bibitem[Brandt et~al\mbox{.}(2018)]%
        {brandt2018modeling}
\bibfield{author}{\bibinfo{person}{Christopher Brandt}, \bibinfo{person}{Leonardo Scandolo}, \bibinfo{person}{Elmar Eisemann}, {and} \bibinfo{person}{Klaus Hildebrandt}.} \bibinfo{year}{2018}\natexlab{}.
\newblock \showarticletitle{Modeling n-symmetry vector fields using higher-order energies}.
\newblock \bibinfo{journal}{\emph{ACM Transactions on Graphics (TOG)}} \bibinfo{volume}{37}, \bibinfo{number}{2} (\bibinfo{year}{2018}), \bibinfo{pages}{1--18}.
\newblock


\bibitem[Campen et~al\mbox{.}(2012)]%
        {campen2012dualloopmeshing}
\bibfield{author}{\bibinfo{person}{Marcel Campen}, \bibinfo{person}{David Bommes}, {and} \bibinfo{person}{Leif Kobbelt}.} \bibinfo{year}{2012}\natexlab{}.
\newblock \showarticletitle{Dual loops meshing: quality quad layouts on manifolds}.
\newblock \bibinfo{journal}{\emph{ACM Transactions on Graphics (TOG)}} \bibinfo{volume}{31}, \bibinfo{number}{4} (\bibinfo{year}{2012}), \bibinfo{pages}{1--11}.
\newblock


\bibitem[Campen and Zorin(2017)]%
        {TSpline2017}
\bibfield{author}{\bibinfo{person}{Marcel Campen} {and} \bibinfo{person}{Denis Zorin}.} \bibinfo{year}{2017}\natexlab{}.
\newblock \showarticletitle{Similarity maps and field-guided T-splines: a perfect couple}.
\newblock \bibinfo{journal}{\emph{ACM Trans. Graph.}} \bibinfo{volume}{36}, \bibinfo{number}{4} (\bibinfo{year}{2017}), \bibinfo{pages}{Article 91}.
\newblock
\showISSN{0730-0301}


\bibitem[Chang et~al\mbox{.}(2015)]%
        {ShapeNet}
\bibfield{author}{\bibinfo{person}{Angel~X. Chang}, \bibinfo{person}{Thomas Funkhouser}, \bibinfo{person}{Leonidas Guibas}, \bibinfo{person}{Pat Hanrahan}, \bibinfo{person}{Qixing Huang}, \bibinfo{person}{Zimo Li}, \bibinfo{person}{Silvio Savarese}, \bibinfo{person}{Manolis Savva}, \bibinfo{person}{Shuran Song}, \bibinfo{person}{Hao Su}, \bibinfo{person}{Jianxiong Xiao}, \bibinfo{person}{Li Yi}, {and} \bibinfo{person}{Fisher Yu}.} \bibinfo{year}{2015}\natexlab{}.
\newblock \bibinfo{booktitle}{\emph{{ShapeNet:} An Information-Rich 3D Model Repository}}.
\newblock \bibinfo{type}{{T}echnical {R}eport} arXiv:1512.03012 [cs.GR].
\newblock


\bibitem[Choy et~al\mbox{.}(2019)]%
        {choy2019sparsecnn}
\bibfield{author}{\bibinfo{person}{Christopher Choy}, \bibinfo{person}{JunYoung Gwak}, {and} \bibinfo{person}{Silvio Savarese}.} \bibinfo{year}{2019}\natexlab{}.
\newblock \showarticletitle{4D Spatio-Temporal ConvNets: Minkowski Convolutional Neural Networks}. In \bibinfo{booktitle}{\emph{Proceedings of the IEEE Conference on Computer Vision and Pattern Recognition}}. \bibinfo{pages}{3075--3084}.
\newblock


\bibitem[Deitke et~al\mbox{.}(2023)]%
        {deitke2023objaverse}
\bibfield{author}{\bibinfo{person}{Matt Deitke}, \bibinfo{person}{Dustin Schwenk}, \bibinfo{person}{Jordi Salvador}, \bibinfo{person}{Luca Weihs}, \bibinfo{person}{Oscar Michel}, \bibinfo{person}{Eli VanderBilt}, \bibinfo{person}{Ludwig Schmidt}, \bibinfo{person}{Kiana Ehsani}, \bibinfo{person}{Aniruddha Kembhavi}, {and} \bibinfo{person}{Ali Farhadi}.} \bibinfo{year}{2023}\natexlab{}.
\newblock \showarticletitle{Objaverse: A universe of annotated 3d objects}. In \bibinfo{booktitle}{\emph{Proceedings of the IEEE/CVF conference on computer vision and pattern recognition}}. \bibinfo{pages}{13142--13153}.
\newblock


\bibitem[Diamanti et~al\mbox{.}(2014)]%
        {PolyVectors_Diamanti2014}
\bibfield{author}{\bibinfo{person}{O. Diamanti}, \bibinfo{person}{A. Vaxman}, \bibinfo{person}{D. Panozzo}, {and} \bibinfo{person}{O. Sorkine-Hornung}.} \bibinfo{year}{2014}\natexlab{}.
\newblock \showarticletitle{Designing n-PolyVector fields with complex polynomials}.
\newblock \bibinfo{journal}{\emph{Computer Graphics Forum}}  \bibinfo{volume}{33} (\bibinfo{year}{2014}), \bibinfo{pages}{1--11}.
\newblock


\bibitem[Diamanti et~al\mbox{.}(2015)]%
        {Integrable2015}
\bibfield{author}{\bibinfo{person}{Olga Diamanti}, \bibinfo{person}{Amir Vaxman}, \bibinfo{person}{Daniele Panozzo}, {and} \bibinfo{person}{Olga Sorkine-Hornung}.} \bibinfo{year}{2015}\natexlab{}.
\newblock \showarticletitle{Integrable {PolyVector} Fields}.
\newblock \bibinfo{journal}{\emph{ACM Transactions on Graphics (proceedings of ACM SIGGRAPH)}} \bibinfo{volume}{34}, \bibinfo{number}{4} (\bibinfo{year}{2015}), \bibinfo{pages}{38:1--38:12}.
\newblock


\bibitem[Dielen et~al\mbox{.}(2021)]%
        {DL2quadMesh2021}
\bibfield{author}{\bibinfo{person}{Alexander Dielen}, \bibinfo{person}{Isaak Lim}, \bibinfo{person}{Max Lyon}, {and} \bibinfo{person}{Leif Kobbelt}.} \bibinfo{year}{2021}\natexlab{}.
\newblock \showarticletitle{Learning direction fields for quad mesh generation}. In \bibinfo{booktitle}{\emph{Computer Graphics Forum}}, Vol.~\bibinfo{volume}{40}. Wiley Online Library, \bibinfo{pages}{181--191}.
\newblock


\bibitem[Dong et~al\mbox{.}(2025)]%
        {NeurCross2024}
\bibfield{author}{\bibinfo{person}{Qiujie Dong}, \bibinfo{person}{Huibiao Wen}, \bibinfo{person}{Rui Xu}, \bibinfo{person}{Shuangmin Chen}, \bibinfo{person}{Jiaran Zhou}, \bibinfo{person}{Shiqing Xin}, \bibinfo{person}{Changhe Tu}, \bibinfo{person}{Taku Komura}, {and} \bibinfo{person}{Wenping Wang}.} \bibinfo{year}{2025}\natexlab{}.
\newblock \showarticletitle{NeurCross: A Neural Approach to Computing Cross Fields for Quad Mesh Generation}.
\newblock \bibinfo{journal}{\emph{ACM Trans. Graph.}} \bibinfo{volume}{44}, \bibinfo{number}{4} (\bibinfo{year}{2025}).
\newblock


\bibitem[Ebke et~al\mbox{.}(2013)]%
        {libQEX13}
\bibfield{author}{\bibinfo{person}{H.-C. Ebke}, \bibinfo{person}{D. Bommes}, \bibinfo{person}{M. Campen}, {and} \bibinfo{person}{L. Kobbelt}.} \bibinfo{year}{2013}\natexlab{}.
\newblock \showarticletitle{QEx: Robust quad mesh extraction}.
\newblock \bibinfo{journal}{\emph{ACM Transactions on Graphics (TOG)}} \bibinfo{volume}{32}, \bibinfo{number}{6} (\bibinfo{year}{2013}), \bibinfo{pages}{168}.
\newblock


\bibitem[Hertzmann and Zorin(2000)]%
        {Illustrating2000}
\bibfield{author}{\bibinfo{person}{Aaron Hertzmann} {and} \bibinfo{person}{Denis Zorin}.} \bibinfo{year}{2000}\natexlab{}.
\newblock \showarticletitle{Illustrating smooth surfaces}. \bibinfo{publisher}{ACM Press/Addison-Wesley Publishing Co.}, \bibinfo{address}{USA}, \bibinfo{pages}{517–526}.
\newblock
\showISBNx{1581132085}


\bibitem[Huang et~al\mbox{.}(2018)]%
        {QuadriFlow2018}
\bibfield{author}{\bibinfo{person}{Jingwei Huang}, \bibinfo{person}{Yichao Zhou}, \bibinfo{person}{Matthias Niessner}, {et~al\mbox{.}}} \bibinfo{year}{2018}\natexlab{}.
\newblock \showarticletitle{QuadriFlow: A Scalable and Robust Method for Quadrangulation}.
\newblock \bibinfo{journal}{\emph{Computer Graphics Forum}} (\bibinfo{year}{2018}).
\newblock
\showISSN{1467-8659}


\bibitem[Jacobson et~al\mbox{.}(2017)]%
        {libigl2017}
\bibfield{author}{\bibinfo{person}{A. Jacobson}, \bibinfo{person}{D. Panozzo}, {and} \bibinfo{person}{et al.}} \bibinfo{year}{2017}\natexlab{}.
\newblock \bibinfo{title}{libigl: A simple C++ geometry processing library}.
\newblock
\newblock
\newblock
\shownote{\url{http://libigl.github.io/libigl/}}.


\bibitem[Jakob et~al\mbox{.}(2015)]%
        {Instant_Meshes2015}
\bibfield{author}{\bibinfo{person}{Wenzel Jakob}, \bibinfo{person}{Marco Tarini}, \bibinfo{person}{Daniele Panozzo}, {and} \bibinfo{person}{Olga Sorkine-Hornung}.} \bibinfo{year}{2015}\natexlab{}.
\newblock \showarticletitle{Instant field-aligned meshes}.
\newblock \bibinfo{journal}{\emph{ACM Transactions on Graphics}} \bibinfo{volume}{34}, \bibinfo{number}{6} (\bibinfo{year}{2015}), \bibinfo{pages}{189}.
\newblock


\bibitem[Kazhdan and Hoppe(2013)]%
        {SPSR2013}
\bibfield{author}{\bibinfo{person}{Michael Kazhdan} {and} \bibinfo{person}{Hugues Hoppe}.} \bibinfo{year}{2013}\natexlab{}.
\newblock \showarticletitle{Screened poisson surface reconstruction}.
\newblock \bibinfo{journal}{\emph{ACM Transactions on Graphics (ToG)}} \bibinfo{volume}{32}, \bibinfo{number}{3} (\bibinfo{year}{2013}), \bibinfo{pages}{1--13}.
\newblock
\showISSN{0730-0301}


\bibitem[Kingma and Jimmy(2014)]%
        {adam2014}
\bibfield{author}{\bibinfo{person}{Diederik~P Kingma} {and} \bibinfo{person}{Ba Jimmy}.} \bibinfo{year}{2014}\natexlab{}.
\newblock \showarticletitle{Adam: A method for stochastic optimization}.
\newblock \bibinfo{journal}{\emph{3rd International Conference for Learning Representations}}.
\newblock


\bibitem[Kingma and Welling(2013)]%
        {kingma2013vae}
\bibfield{author}{\bibinfo{person}{Diederik~P Kingma} {and} \bibinfo{person}{Max Welling}.} \bibinfo{year}{2013}\natexlab{}.
\newblock \bibinfo{title}{Auto-encoding variational Bayes}.
\newblock \bibinfo{howpublished}{arXiv:1312.6114}.
\newblock


\bibitem[Knöppel et~al\mbox{.}(2013)]%
        {Power_Fields2013}
\bibfield{author}{\bibinfo{person}{Felix Knöppel}, \bibinfo{person}{Keenan Crane}, \bibinfo{person}{Ulrich Pinkall}, {and} \bibinfo{person}{Peter Schröder}.} \bibinfo{year}{2013}\natexlab{}.
\newblock \showarticletitle{Globally Optimal Direction Fields}.
\newblock \bibinfo{journal}{\emph{ACM Transactions on Graphics}} \bibinfo{volume}{32}, \bibinfo{number}{4} (\bibinfo{year}{2013}), \bibinfo{pages}{1--10}.
\newblock


\bibitem[Koch et~al\mbox{.}(2019)]%
        {Koch_2019_abc}
\bibfield{author}{\bibinfo{person}{Sebastian Koch}, \bibinfo{person}{Albert Matveev}, \bibinfo{person}{Zhongshi Jiang}, \bibinfo{person}{Francis Williams}, \bibinfo{person}{Alexey Artemov}, \bibinfo{person}{Evgeny Burnaev}, \bibinfo{person}{Marc Alexa}, \bibinfo{person}{Denis Zorin}, {and} \bibinfo{person}{Daniele Panozzo}.} \bibinfo{year}{2019}\natexlab{}.
\newblock \showarticletitle{ABC: A Big CAD Model Dataset For Geometric Deep Learning}. In \bibinfo{booktitle}{\emph{The IEEE Conference on Computer Vision and Pattern Recognition (CVPR)}}.
\newblock


\bibitem[Li et~al\mbox{.}(2021)]%
        {li20214deform4d}
\bibfield{author}{\bibinfo{person}{Yang Li}, \bibinfo{person}{Hikari Takehara}, \bibinfo{person}{Takafumi Taketomi}, \bibinfo{person}{Bo Zheng}, {and} \bibinfo{person}{Matthias Nie{\ss}ner}.} \bibinfo{year}{2021}\natexlab{}.
\newblock \showarticletitle{4dcomplete: Non-rigid motion estimation beyond the observable surface}. In \bibinfo{booktitle}{\emph{Proceedings of the IEEE/CVF International Conference on Computer Vision}}. \bibinfo{pages}{12706--12716}.
\newblock


\bibitem[Li et~al\mbox{.}(2025)]%
        {li2025point2quad}
\bibfield{author}{\bibinfo{person}{Zezeng Li}, \bibinfo{person}{Zhihui Qi}, \bibinfo{person}{Weimin Wang}, \bibinfo{person}{Ziliang Wang}, \bibinfo{person}{Junyi Duan}, {and} \bibinfo{person}{Na Lei}.} \bibinfo{year}{2025}\natexlab{}.
\newblock \showarticletitle{Point2Quad: Generating Quad Meshes from Point Clouds via Face Prediction}.
\newblock \bibinfo{journal}{\emph{IEEE Transactions on Circuits and Systems for Video Technology}} (\bibinfo{year}{2025}).
\newblock


\bibitem[Loop and Schaefer(2008)]%
        {CC_subdivision2008}
\bibfield{author}{\bibinfo{person}{Charles Loop} {and} \bibinfo{person}{Scott Schaefer}.} \bibinfo{year}{2008}\natexlab{}.
\newblock \showarticletitle{Approximating Catmull-Clark subdivision surfaces with bicubic patches}.
\newblock \bibinfo{journal}{\emph{ACM Trans. Graph.}} \bibinfo{volume}{27}, \bibinfo{number}{1}, Article \bibinfo{articleno}{8} (\bibinfo{date}{March} \bibinfo{year}{2008}), \bibinfo{numpages}{11}~pages.
\newblock
\showISSN{0730-0301}


\bibitem[Lorensen and Cline(1987)]%
        {MC}
\bibfield{author}{\bibinfo{person}{William~E. Lorensen} {and} \bibinfo{person}{Harvey~E. Cline}.} \bibinfo{year}{1987}\natexlab{}.
\newblock \showarticletitle{Marching Cubes: A High Resolution 3D Surface Construction Algorithm}. In \bibinfo{booktitle}{\emph{SIGGRAPH}}.
\newblock


\bibitem[Mittal et~al\mbox{.}(2022)]%
        {mittal2022autosdf}
\bibfield{author}{\bibinfo{person}{Paritosh Mittal}, \bibinfo{person}{Yen-Chi Cheng}, \bibinfo{person}{Maneesh Singh}, {and} \bibinfo{person}{Shubham Tulsiani}.} \bibinfo{year}{2022}\natexlab{}.
\newblock \showarticletitle{AutoSDF: Shape priors for 3D completion, reconstruction and generation}. In \bibinfo{booktitle}{\emph{Proceedings of the IEEE/CVF Conference on Computer Vision and Pattern Recognition}}. \bibinfo{pages}{306--315}.
\newblock


\bibitem[Mo et~al\mbox{.}(2019)]%
        {mo2019structurenet}
\bibfield{author}{\bibinfo{person}{Kaichun Mo}, \bibinfo{person}{Paul Guerrero}, \bibinfo{person}{Li Yi}, \bibinfo{person}{Hao Su}, \bibinfo{person}{Peter Wonka}, \bibinfo{person}{Niloy Mitra}, {and} \bibinfo{person}{Leonidas Guibas}.} \bibinfo{year}{2019}\natexlab{}.
\newblock \showarticletitle{StructureNet: Hierarchical Graph Networks for 3D Shape Generation}.
\newblock \bibinfo{journal}{\emph{ACM Transactions on Graphics (TOG), Siggraph Asia 2019}} \bibinfo{volume}{38}, \bibinfo{number}{6} (\bibinfo{year}{2019}), \bibinfo{pages}{Article 242}.
\newblock


\bibitem[Palmer et~al\mbox{.}(2024)]%
        {palmer2024lifting}
\bibfield{author}{\bibinfo{person}{David Palmer}, \bibinfo{person}{Albert Chern}, {and} \bibinfo{person}{Justin Solomon}.} \bibinfo{year}{2024}\natexlab{}.
\newblock \showarticletitle{Lifting Directional Fields to Minimal Sections}.
\newblock \bibinfo{journal}{\emph{ACM Transactions on Graphics (TOG)}} \bibinfo{volume}{43}, \bibinfo{number}{4} (\bibinfo{year}{2024}), \bibinfo{pages}{1--20}.
\newblock


\bibitem[Panozzo et~al\mbox{.}(2014)]%
        {Panozzo2014}
\bibfield{author}{\bibinfo{person}{Daniele Panozzo}, \bibinfo{person}{Enrico Puppo}, \bibinfo{person}{Marco Tarini}, {and} \bibinfo{person}{Olga Sorkine-Hornung}.} \bibinfo{year}{2014}\natexlab{}.
\newblock \showarticletitle{Frame Fields: Anisotropic and Non-Orthogonal Cross Fields}.
\newblock \bibinfo{journal}{\emph{ACM Transactions on Graphics (proceedings of ACM SIGGRAPH)}} \bibinfo{volume}{33}, \bibinfo{number}{4} (\bibinfo{year}{2014}), \bibinfo{pages}{134:1--134:11}.
\newblock


\bibitem[Pietroni et~al\mbox{.}(2021)]%
        {quadwild2021}
\bibfield{author}{\bibinfo{person}{Nico Pietroni}, \bibinfo{person}{Stefano Nuvoli}, \bibinfo{person}{Thomas Alderighi}, \bibinfo{person}{Paolo Cignoni}, {and} \bibinfo{person}{Marco Tarini}.} \bibinfo{year}{2021}\natexlab{}.
\newblock \showarticletitle{Reliable feature-line driven quad-remeshing}.
\newblock \bibinfo{journal}{\emph{ACM Trans. Graph.}} \bibinfo{volume}{40}, \bibinfo{number}{4} (\bibinfo{year}{2021}), \bibinfo{pages}{Article 155}.
\newblock
\showISSN{0730-0301}
\urldef\tempurl%
\url{https://doi.org/10.1145/3450626.3459941}
\showDOI{\tempurl}


\bibitem[Qi et~al\mbox{.}(2017)]%
        {QSMG17}
\bibfield{author}{\bibinfo{person}{Charles~R Qi}, \bibinfo{person}{Hao Su}, \bibinfo{person}{Kaichun Mo}, {and} \bibinfo{person}{Leonidas~J. Guibas}.} \bibinfo{year}{2017}\natexlab{}.
\newblock \showarticletitle{PointNet: Deep Learning on Point Sets for 3D Classification and Segmentation}. In \bibinfo{booktitle}{\emph{IEEE Conf. on Computer Vision and Pattern Recognition}}. \bibinfo{pages}{652--660}.
\newblock


\bibitem[Ray et~al\mbox{.}(2008)]%
        {N-symmetry2008}
\bibfield{author}{\bibinfo{person}{Nicolas Ray}, \bibinfo{person}{Bruno Vallet}, \bibinfo{person}{Wan~Chiu Li}, {and} \bibinfo{person}{Bruno L\'{e}vy}.} \bibinfo{year}{2008}\natexlab{}.
\newblock \showarticletitle{N-symmetry direction field design}.
\newblock \bibinfo{journal}{\emph{ACM Trans. Graph.}} \bibinfo{volume}{27}, \bibinfo{number}{2}, Article \bibinfo{articleno}{10} (\bibinfo{year}{2008}), \bibinfo{numpages}{13}~pages.
\newblock
\showISSN{0730-0301}


\bibitem[Rombach et~al\mbox{.}(2022)]%
        {rombach2022stablediffusion}
\bibfield{author}{\bibinfo{person}{Robin Rombach}, \bibinfo{person}{Andreas Blattmann}, \bibinfo{person}{Dominik Lorenz}, \bibinfo{person}{Patrick Esser}, {and} \bibinfo{person}{Bj{\"o}rn Ommer}.} \bibinfo{year}{2022}\natexlab{}.
\newblock \showarticletitle{High-resolution image synthesis with latent diffusion models}. In \bibinfo{booktitle}{\emph{Proceedings of the IEEE/CVF Conference on Computer Vision and Pattern Recognition (CVPR)}}.
\newblock


\bibitem[Sageman-Furnas et~al\mbox{.}(2019)]%
        {Chebyshev2019}
\bibfield{author}{\bibinfo{person}{Andrew~O. Sageman-Furnas}, \bibinfo{person}{Albert Chern}, \bibinfo{person}{Mirela Ben-Chen}, {and} \bibinfo{person}{Amir Vaxman}.} \bibinfo{year}{2019}\natexlab{}.
\newblock \showarticletitle{Chebyshev nets from commuting PolyVector fields}.
\newblock \bibinfo{journal}{\emph{ACM Trans. Graph.}} \bibinfo{volume}{38}, \bibinfo{number}{6}, Article \bibinfo{articleno}{172} (\bibinfo{date}{Nov.} \bibinfo{year}{2019}), \bibinfo{numpages}{16}~pages.
\newblock
\showISSN{0730-0301}


\bibitem[Shen et~al\mbox{.}(2014)]%
        {subdivision2014}
\bibfield{author}{\bibinfo{person}{Jingjing Shen}, \bibinfo{person}{Jiří Kosinka}, \bibinfo{person}{Malcolm~A. Sabin}, {and} \bibinfo{person}{Neil~A. Dodgson}.} \bibinfo{year}{2014}\natexlab{}.
\newblock \showarticletitle{Conversion of trimmed NURBS surfaces to Catmull–Clark subdivision surfaces}.
\newblock \bibinfo{journal}{\emph{Computer Aided Geometric Design}} \bibinfo{volume}{31}, \bibinfo{number}{7} (\bibinfo{year}{2014}), \bibinfo{pages}{486--498}.
\newblock
\showISSN{0167-8396}
\newblock
\shownote{Recent Trends in Theoretical and Applied Geometry}.


\bibitem[Van Den~Oord et~al\mbox{.}(2017)]%
        {van2017vqvae}
\bibfield{author}{\bibinfo{person}{Aaron Van Den~Oord}, \bibinfo{person}{Oriol Vinyals}, {et~al\mbox{.}}} \bibinfo{year}{2017}\natexlab{}.
\newblock \showarticletitle{Neural discrete representation learning}.
\newblock \bibinfo{journal}{\emph{Advances in neural information processing systems}}  \bibinfo{volume}{30} (\bibinfo{year}{2017}).
\newblock


\bibitem[Vaxman et~al\mbox{.}(2016)]%
        {fieldDesign2016}
\bibfield{author}{\bibinfo{person}{Amir Vaxman}, \bibinfo{person}{Marcel Campen}, \bibinfo{person}{Olga Diamanti}, {et~al\mbox{.}}} \bibinfo{year}{2016}\natexlab{}.
\newblock \showarticletitle{Directional Field Synthesis, Design, and Processing}.
\newblock \bibinfo{journal}{\emph{Computer Graphics Forum}} \bibinfo{volume}{35}, \bibinfo{number}{2} (\bibinfo{year}{2016}), \bibinfo{pages}{545--572}.
\newblock


\bibitem[Vaxman et~al\mbox{.}(2017)]%
        {survery2017}
\bibfield{author}{\bibinfo{person}{Amir Vaxman}, \bibinfo{person}{Marcel Campen}, \bibinfo{person}{Olga Diamanti}, \bibinfo{person}{David Bommes}, \bibinfo{person}{Klaus Hildebrandt}, \bibinfo{person}{Mirela Ben-Chen Technion}, {and} \bibinfo{person}{Daniele Panozzo}.} \bibinfo{year}{2017}\natexlab{}.
\newblock \showarticletitle{Directional field synthesis, design, and processing}. In \bibinfo{booktitle}{\emph{ACM SIGGRAPH 2017 Courses}} (Los Angeles, California) \emph{(\bibinfo{series}{SIGGRAPH '17})}. \bibinfo{publisher}{Association for Computing Machinery}, \bibinfo{address}{New York, NY, USA}, Article \bibinfo{articleno}{12}, \bibinfo{numpages}{30}~pages.
\newblock
\showISBNx{9781450350143}


\bibitem[Viertel and Osting(2019)]%
        {viertel2019approach}
\bibfield{author}{\bibinfo{person}{Ryan Viertel} {and} \bibinfo{person}{Braxton Osting}.} \bibinfo{year}{2019}\natexlab{}.
\newblock \showarticletitle{An approach to quad meshing based on harmonic cross-valued maps and the Ginzburg--Landau theory}.
\newblock \bibinfo{journal}{\emph{SIAM Journal on Scientific Computing}} \bibinfo{volume}{41}, \bibinfo{number}{1} (\bibinfo{year}{2019}), \bibinfo{pages}{A452--A479}.
\newblock


\bibitem[Vouga et~al\mbox{.}(2012)]%
        {architecturalGeometry2012}
\bibfield{author}{\bibinfo{person}{Etienne Vouga}, \bibinfo{person}{Mathias H\"{o}binger}, \bibinfo{person}{Johannes Wallner}, {and} \bibinfo{person}{Helmut Pottmann}.} \bibinfo{year}{2012}\natexlab{}.
\newblock \showarticletitle{Design of self-supporting surfaces}.
\newblock \bibinfo{journal}{\emph{ACM Trans. Graph.}} \bibinfo{volume}{31}, \bibinfo{number}{4}, Article \bibinfo{articleno}{87} (\bibinfo{date}{jul} \bibinfo{year}{2012}), \bibinfo{numpages}{11}~pages.
\newblock
\showISSN{0730-0301}


\bibitem[Wang et~al\mbox{.}(2025)]%
        {StructRe2025}
\bibfield{author}{\bibinfo{person}{Jiepeng Wang}, \bibinfo{person}{Hao Pan}, \bibinfo{person}{Yang Liu}, \bibinfo{person}{Xin Tong}, \bibinfo{person}{Taku Komura}, {and} \bibinfo{person}{Wenping Wang}.} \bibinfo{year}{2025}\natexlab{}.
\newblock \showarticletitle{StructRe: Rewriting for Structured Shape Modeling}.
\newblock \bibinfo{journal}{\emph{ACM Trans. Graph.}} (\bibinfo{date}{April} \bibinfo{year}{2025}).
\newblock
\showISSN{0730-0301}
\urldef\tempurl%
\url{https://doi.org/10.1145/3732934}
\showDOI{\tempurl}


\bibitem[Werner et~al\mbox{.}(2014)]%
        {TSDF}
\bibfield{author}{\bibinfo{person}{Diana Werner}, \bibinfo{person}{Ayoub Al-Hamadi}, {and} \bibinfo{person}{Philipp Werner}.} \bibinfo{year}{2014}\natexlab{}.
\newblock \showarticletitle{Truncated Signed Distance Function: Experiments on Voxel Size}. In \bibinfo{booktitle}{\emph{Image Analysis and Recognition}}, \bibfield{editor}{\bibinfo{person}{Aur{\'e}lio Campilho} {and} \bibinfo{person}{Mohamed Kamel}} (Eds.). \bibinfo{publisher}{Springer International Publishing}, \bibinfo{address}{Cham}, \bibinfo{pages}{357--364}.
\newblock
\showISBNx{978-3-319-11755-3}


\bibitem[Xiang et~al\mbox{.}(2024)]%
        {xiang2024TRELLIS}
\bibfield{author}{\bibinfo{person}{Jianfeng Xiang}, \bibinfo{person}{Zelong Lv}, \bibinfo{person}{Sicheng Xu}, \bibinfo{person}{Yu Deng}, \bibinfo{person}{Ruicheng Wang}, \bibinfo{person}{Bowen Zhang}, \bibinfo{person}{Dong Chen}, \bibinfo{person}{Xin Tong}, {and} \bibinfo{person}{Jiaolong Yang}.} \bibinfo{year}{2024}\natexlab{}.
\newblock \showarticletitle{Structured 3D Latents for Scalable and Versatile 3D Generation}.
\newblock \bibinfo{journal}{\emph{arXiv preprint arXiv:2412.01506}} (\bibinfo{year}{2024}).
\newblock


\bibitem[Xu et~al\mbox{.}(2023)]%
        {normal_Xu506}
\bibfield{author}{\bibinfo{person}{Rui Xu}, \bibinfo{person}{Zhiyang Dou}, \bibinfo{person}{Ningna Wang}, \bibinfo{person}{Shiqing Xin}, \bibinfo{person}{Shuangmin Chen}, \bibinfo{person}{Mingyan Jiang}, \bibinfo{person}{Xiaohu Guo}, \bibinfo{person}{Wenping Wang}, {and} \bibinfo{person}{Changhe Tu}.} \bibinfo{year}{2023}\natexlab{}.
\newblock \showarticletitle{Globally Consistent Normal Orientation for Point Clouds by Regularizing the Winding-Number Field}.
\newblock \bibinfo{journal}{\emph{ACM Trans. Graph.}} \bibinfo{volume}{42}, \bibinfo{number}{4} (\bibinfo{year}{2023}), \bibinfo{pages}{Article 111}.
\newblock
\showISSN{0730-0301}
\urldef\tempurl%
\url{https://doi.org/10.1145/3592129}
\showDOI{\tempurl}


\bibitem[Yan et~al\mbox{.}(2022)]%
        {yan2022shapeformer}
\bibfield{author}{\bibinfo{person}{Xingguang Yan}, \bibinfo{person}{Liqiang Lin}, \bibinfo{person}{Niloy~J Mitra}, \bibinfo{person}{Dani Lischinski}, \bibinfo{person}{Daniel Cohen-Or}, {and} \bibinfo{person}{Hui Huang}.} \bibinfo{year}{2022}\natexlab{}.
\newblock \showarticletitle{Shapeformer: Transformer-based shape completion via sparse representation}. In \bibinfo{booktitle}{\emph{IEEE/CVF Conference on Computer Vision and Pattern Recognition}}.
\newblock


\bibitem[Zhang et~al\mbox{.}(2020)]%
        {Justin2020}
\bibfield{author}{\bibinfo{person}{Paul Zhang}, \bibinfo{person}{Josh Vekhter}, \bibinfo{person}{Edward Chien}, \bibinfo{person}{David Bommes}, \bibinfo{person}{Etienne Vouga}, {and} \bibinfo{person}{Justin Solomon}.} \bibinfo{year}{2020}\natexlab{}.
\newblock \showarticletitle{Octahedral Frames for Feature-Aligned Cross Fields}.
\newblock \bibinfo{journal}{\emph{ACM Trans. Graph.}} \bibinfo{volume}{39}, \bibinfo{number}{3}, Article \bibinfo{articleno}{25} (\bibinfo{date}{April} \bibinfo{year}{2020}), \bibinfo{numpages}{13}~pages.
\newblock
\showISSN{0730-0301}


\bibitem[Zheng et~al\mbox{.}(2023)]%
        {zheng2023LASDiffusion}
\bibfield{author}{\bibinfo{person}{Xin-Yang Zheng}, \bibinfo{person}{Hao Pan}, \bibinfo{person}{Peng-Shuai Wang}, \bibinfo{person}{Xin Tong}, \bibinfo{person}{Yang Liu}, {and} \bibinfo{person}{Heung-Yeung Shum}.} \bibinfo{year}{2023}\natexlab{}.
\newblock \showarticletitle{Locally Attentional SDF Diffusion for Controllable 3D Shape Generation}.
\newblock \bibinfo{journal}{\emph{ACM Transactions on Graphics (SIGGRAPH)}} \bibinfo{volume}{42}, \bibinfo{number}{4} (\bibinfo{year}{2023}).
\newblock


\bibitem[Zhou and Jacobson(2016)]%
        {Thingi10K}
\bibfield{author}{\bibinfo{person}{Qingnan Zhou} {and} \bibinfo{person}{Alec Jacobson}.} \bibinfo{year}{2016}\natexlab{}.
\newblock \showarticletitle{Thingi10K: A Dataset of 10,000 3D-Printing Models}.
\newblock \bibinfo{journal}{\emph{arXiv preprint arXiv:1605.04797}} (\bibinfo{year}{2016}).
\newblock


\end{thebibliography}

\clearpage
\begin{figure*}[t]
  \centering
  \begin{overpic}[width=\linewidth]{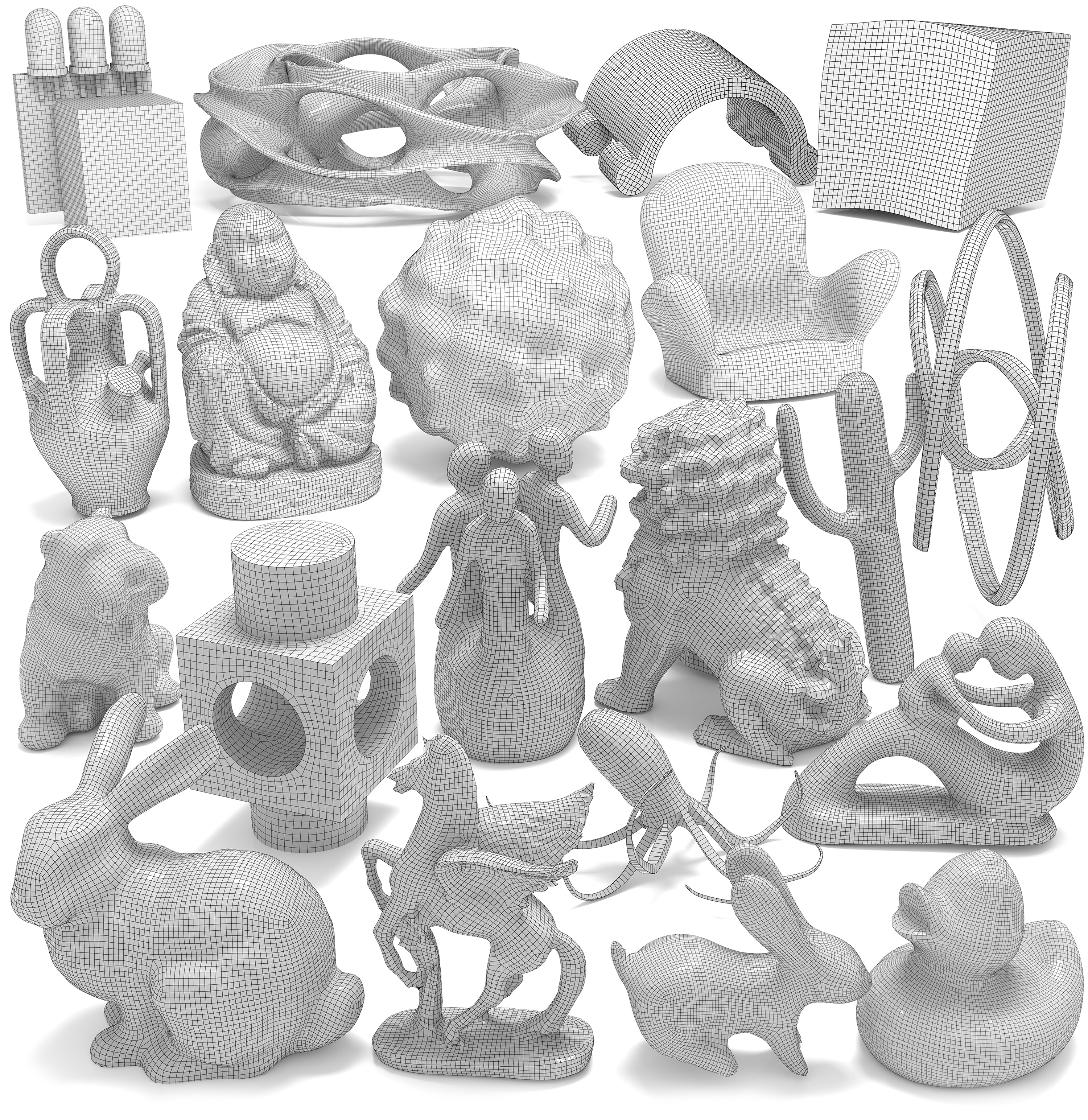}
  \end{overpic}
   \caption{
   Gallery of cross field generation for quad meshing. We evaluate our method on a variety of shapes, including out-of-domain examples, for feed-forward cross field generation and visualize the quad meshes extracted from the cross fields. 
   }
   \label{fig:gallery_quad}
\end{figure*}

\end{document}